\documentclass{emulateapj}

\renewcommand{\ng}{\bar{n}_\mathrm{g}}
\newcommand{\Ng}{N_\mathrm{g}}
\newcommand{\Lbox}{L_{\mathrm{box}}}
\newcommand{\zmin}{z_{\mathrm{min}}}
\newcommand{\zmax}{z_{\mathrm{max}}}
\newcommand{\hden}{h^{3}{\mathrm{Mpc}}^{-3}}
\newcommand{\hvol}{h^{-3}{\mathrm{Mpc}}^{3}}
\newcommand{\hmpc}{h^{-1}\mathrm{Mpc}}
\newcommand{\hkpc}{h^{-1}\mathrm{kpc}}
\newcommand{\hMsun}{h^{-1}M_{\odot}}

\newcommand{\kms}{{\,{\rm km}\,{\rm s}^{-1}}}
\newcommand{\Omegam}{\Omega_{m}}
\newcommand{\Omegab}{\Omega_{b}}
\newcommand{\Omegal}{\Omega_{\Lambda}}
\newcommand{\Mmin}{M_\mathrm{min}}
\newcommand{\Mcut}{M_\mathrm{cut}}

\newcommand{\Nsat}{\left<N_{\mathrm{sat}}\right>}
\newcommand{\band}[2]{\ensuremath{^{#1}\!{#2}}}
\newcommand{\Mr}{M_{\band{0.1}{r}}}
\newcommand{\gr}{\band{0.1}{(g-r)}}
\newcommand{\Mgtot}{M_{g20}}
\newcommand{\Mrtot}{M_{r20}}
\newcommand{\Lrtot}{L_{r20}}
\newcommand{\Lstar}{L_\ast}
\newcommand{\Mstar}{M_\ast}
\newcommand{\grgrp}{(g-r)_{20}}
\newcommand{\sigv}{\sigma_v}
\newcommand{\Rproj}{R_{\perp,\mathrm{rms}}}
\newcommand{\redge}{r_\mathrm{edge}}
\newcommand{\ngrp}{n_\mathrm{grp}}
\newcommand{\ngrpN}{n_\mathrm{grp}(N)}
\newcommand{\Nmin}{N_{\mathrm{min}}}
\newcommand{\Nmax}{N_{\mathrm{max}}}
\newcommand{\signgrpN}{\sigma_{n_\mathrm{grp}}}

\newcommand{\Rrms}{R_\mathrm{rms}}
\newcommand{\bpar}{b_{\parallel}}
\newcommand{\bperp}{b_{\perp}}
\newcommand{\Ntrue}{N_\mathrm{true}}
\newcommand{\Nobs}{N_\mathrm{obs}}
\newcommand{\Hunits}{{\,{\rm km}\,{\rm s}^{-1}\,{\rm Mpc}^{-1}}}
\newcommand{\tcross}{t_\mathrm{cross}}
\newcommand{\tH}{t_\mathrm{H}}

\bibpunct[,]{(}{)}{;}{a}{}{,}
\begin{document}

\title{Percolation Galaxy Groups and Clusters in the SDSS Redshift Survey: Identification, Catalogs, and the Multiplicity Function}

\author{
Andreas A. Berlind, \altaffilmark{1,2}
Joshua Frieman, \altaffilmark{2}
David H. Weinberg, \altaffilmark{3}
Michael R. Blanton, \altaffilmark{1}
Michael S. Warren, \altaffilmark{4}
Kevork Abazajian, \altaffilmark{4}
Ryan Scranton, \altaffilmark{5}
David W. Hogg, \altaffilmark{1}
Roman Scoccimarro, \altaffilmark{1}
Neta A. Bahcall, \altaffilmark{6}
J. Brinkmann, \altaffilmark{7}
J. Richard Gott III, \altaffilmark{6}
S.J. Kleinman, \altaffilmark{7}
J. Krzesinski, \altaffilmark{8,9}
Brian C. Lee, \altaffilmark{10}
Christopher J. Miller, \altaffilmark{11}
Atsuko Nitta, \altaffilmark{7}
Donald P. Schneider, \altaffilmark{12}
Douglas L. Tucker, \altaffilmark{13}
Idit Zehavi, \altaffilmark{14,15}
for the SDSS collaboration
}

\begin{abstract}
We identify galaxy groups and clusters in volume-limited samples of the SDSS
redshift survey, using a redshift-space friends-of-friends algorithm.  We optimize the
friends-of-friends linking lengths to recover galaxy systems that occupy the same 
dark matter halos, using a set of mock catalogs created by populating halos of 
N-body simulations with galaxies.  Extensive tests with these mock catalogs show 
that no combination of perpendicular and line-of-sight linking lengths is able to 
yield groups and clusters that simultaneously recover the true halo multiplicity 
function, projected size distribution, and velocity dispersion.  We adopt a linking 
length combination that yields, for galaxy groups with ten or more members: a group 
multiplicity function that is unbiased with respect to the true halo multiplicity
function; an unbiased median relation between the multiplicities of groups
and their associated halos; a spurious group fraction of less than $\sim 1\%$;
a halo completeness of more than $\sim 97\%$; the correct projected size
distribution as a function of multiplicity; and a velocity dispersion distribution
that is $\sim 20\%$ too low at all multiplicities.  These results hold over
a range of mock catalogs that use different input recipes of populating halos
with galaxies.  We apply our group-finding algorithm to the SDSS data and obtain
three group and cluster catalogs for three volume-limited samples that cover 3495.1
square degrees on the sky, go out to redshifts of 0.1, 0.068, and 0.045, and contain
57138, 37820, and 18895 galaxies, respectively.  We correct for incompleteness caused
by fiber collisions and survey edges, and obtain measurements of the group
multiplicity function, with errors calculated from realistic mock catalogs.
These multiplicity function measurements provide a key constraint on the relation 
between galaxy populations and dark matter halos.
\end{abstract}


\altaffiltext{1}{Center for Cosmology and Particle Physics, New York 
University, New York, NY 10003, USA; aberlind@cosmo.nyu.edu}
\altaffiltext{2}{Center for Cosmological Physics and Department of Astronomy
and Astrophysics, University of Chicago, Chicago, IL 60637; frieman@fnal.gov}
\altaffiltext{3}{Department of Astronomy, The Ohio State University, Columbus, 
OH 43210; dhw@astronomy.ohio-state.edu}
\altaffiltext{4}{Theoretical Division, Los Alamos National Laboratory, Los Alamos,
NM 87545}
\altaffiltext{5}{Physics and Astronomy Department, University of Pittsburgh,
Pittsburgh PA, 15260}
\altaffiltext{6}{Department of Astrophysical Sciences, Princeton University,
Princeton NJ, 08544}
\altaffiltext{7}{Subaru Telescope, 650 N A'ohoku Pl., Hilo, HI 96720}
\altaffiltext{8}{Apache Point Observatory, P.O. Box 59, Sunspot, NM 88349}
\altaffiltext{9}{Mt. Suhora Observatory, Cracow Pedagogical University, ul. 
Podchorazych 2, 30-084 Cracow, Poland}
\altaffiltext{10}{Lawrence Berkeley National Lab, Berkeley CA 94720}
\altaffiltext{11}{Cerro-Tololo Inter-American Observatory, NOAO, Casilla 603, La Serena, 
Chile}
\altaffiltext{12}{Department of Astronomy and Astrophysics, Pennsylvania State University,
University Park, PA 16802}
\altaffiltext{13}{Fermi National Accelerator Laboratory, MS 127, PO Box 500, 
Batavia, IL 60510}

\altaffiltext{14}{Steward Observatory, University of Arizona, 933 N. Cherry Ave., 
Tucson AZ 85721}
\altaffiltext{15}{Deptartment of Astronomy, Case Western Reserve University, 
Cleveland, OH 44106}
%#############################################################################

\section{Introduction} \label{intro}

Galaxies are gregarious by nature.  Bright galaxies typically reside in groups
or clusters, surrounded by less luminous neighbors.  Interactions within the
group or cluster environment may have important effects on the star formation
history, morphology, dynamics, and other properties of member galaxies.
Characterizing the relation between galaxy properties and their group environment
is thus a key step in understanding galaxy formation and evolution.
At the density thresholds often used to identify groups, most members should
belong to the same, gravitationally bound dark matter (DM) halo.\footnote{Throughout
this paper, we use the term ``halo'' to refer to a gravitationally bound structure
with overdensity $\rho/\bar{\rho}\sim200$, so an occupied halo may host a single
luminous galaxy, a group of galaxies, or a cluster.  Higher overdensity concentrations
around individual galaxies of a group or cluster constitute, in this terminology,
halo substructure, or ``sub-halos''.}  Recent approaches to describing the relation 
between galaxies and DM focus on galaxy populations of DM halos as a function of
halo mass.  Specifically, the bias of a particular class of galaxies can be
characterized by its Halo Occupation Distribution (HOD), which specifies the
probability distribution $P(N|M)$ that a halo of mass $M$ contains $N$ such galaxies,
together with relations describing the relative spatial and velocity distributions
of galaxies and dark matter within halos (\citealt{berlind_weinberg_02} and references 
therein).  A well defined group catalog with well understood properties can play
a central role in the empirical determination of this relation.

This paper presents a group and cluster catalog defined from the Sloan Digital Sky
Survey (SDSS, \citealt{york_etal_00}).  While this catalog is useful for many purposes, 
our overriding objective is to obtain a well understood measurement of the group 
multiplicity function (the space density of groups as a function of richness), with the 
goal of determining the HOD in the high mass regime \citep{peacock_smith_00,berlind_weinberg_02,marinoni_hudson_02,kochanek_etal_03,lin_etal_04}.
With this objective in mind, we have adopted a simple group-finding algorithm, 
friends-of-friends in redshift space \citep{huchra_geller_82}, and carried out 
extensive tests on realistic mock catalogs in order to assess its performance and 
optimize parameter choices.  We apply the group-finding algorithm to volume-limited 
samples of galaxies so that the resulting group statistics characterize the clustering 
of well defined populations of galaxies.

Galaxy clusters have been the focus of study since they were first seen on optical 
photographic plates \citep{shapley_ames_26}.  \citet{zwicky_37} pioneered the study
of clusters as dynamical objects by using imaging and spectroscopy of the Coma cluster
to estimate its mass.  However, the most influential pioneering work on clusters was 
done by \citet{abell_58}, who assembled the first large sample of galaxy clusters.  The 
Abell catalog of rich galaxy clusters \citep{abell_58,abell_etal_89} was created by 
eyeball identification in the Palomar Observatory Sky Survey and it spawned numerous 
follow-up studies.  \citet{devaucouleurs_71} shifted focus to poorer systems by 
studying nearby groups of galaxies.  \citet{gott_turner_77b} made the first measurement 
of the group multiplicity function using the \citep{turner_gott_76} catalog of groups
selected based on the projected surface density of galaxies.

With the advent of large redshift surveys, group identification became three dimensional
and thus less subject to projection effects.  Group-finding in redshift space was
pioneered by \citet{huchra_geller_82} and \citet{geller_huchra_83}, using the Center for
Astrophysics (CfA) redshift survey.  Subsequent versions of the CfA redshift survey were 
used to identify groups by various authors 
\citep{nolthenius_white_87,ramella_etal_89,moore_etal_93,ramella_etal_97}.
Other redshift surveys that spawned group catalogs were the Nearby Galaxies Catalog 
\citep{tully_87}, the ESO Slice Project \citep{ramella_etal_99}, the Las Campanas
Redshift Survey (LCRS) \citep{tucker_etal_00}, the Nearby Optical Galaxy Sample (NOG) 
\citep{giuricin_etal_00}, the Southern Sky Redshift Survey (SSRS) 
\citep{ramella_etal_02}, the 2dF redshift survey 
\citep{merchan_zandivarez_02,eke_etal_04a,yang_etal_05}, and even the high redshift DEEP2
survey \citep{gerke_etal_05}.

There have been several efforts to detect clusters in the SDSS to date, most of them
using the photometric data rather than the redshift data.  \citet{annis_etal_99}
developed the maxBCG technique, where Brightest Cluster Galaxy (BCG) candidates 
are identified based on their colors and magnitudes and other cluster members are 
selected from nearby galaxies that have the colors of the E/S0 ridgeline.  
\citet{kim_etal_02} developed a hybrid matched filter (HMF) technique that assumes
a radial profile for clusters and convolves the data with that filter.  
\citet{goto_etal_02} developed the cut-and-enhance (CE) method, which selects 
overdensities of galaxies that have similar colors.  All these techniques were applied 
to the early SDSS commissioning data \citep{bahcall_etal_03,goto_etal_02}.  
\citet{lee_etal_04} identified compact groups by looking for small and isolated
concentrations of galaxies in the SDSS Early Data Release (EDR;
\citealt{stoughton_etal_02}).  Cluster searches in the SDSS redshift survey have also
been carried out.  \citet{goto_etal_05} used a friends-of-friends algorithm (though
with linking lengths that do not scale with the changing number density of galaxies
due to the flux limit) to identify clusters in the SDSS Data Release 2 (DR2; 
\citealt{abazajian_etal_04}).  \citet{merchan_zandivarez_05} used a friends-of-friends
algorithm to identify groups in the SDSS Data Release 3 (DR3; 
\citealt{abazajian_etal_05}).  \citet{weinmann_etal_05} used the \citet{yang_etal_05}
algorithm to identify groups in SDSS DR2.  \citet{miller_etal_05} developed the C4 
algorithm for finding clusters in redshift space and also applied it to the SDSS DR2.  
The C4 algorithm looks for concentrations of galaxies in a seven-dimensional position 
and color space.  It takes advantage of the color similarity of cluster member galaxies 
and thus minimizes contamination due to projection.  However, some correlations are built
into the method, and modeling it in order to understand the properties of the resulting 
cluster catalog requires a complete model of the galaxy population (including colors
and luminosities).  Our method complements the C4 catalog by applying a simple and 
easily modeled algorithm to volume-limited samples with homogeneous properties.

In \S~\ref{data} we describe the SDSS data that we use.  In \S~\ref{mocks} we describe
the mock catalogs that we use to optimize our group-finder and to estimate uncertainties
for our measured group statistics.  In \S~\ref{groupfinder} we outline our group-finding
algorithm and choice of parameters.  We present a detailed discussion of tests with mock 
catalogs in the Appendix, with the key points summarized in the main text.
We discuss incompleteness in our group catalogs due to fiber collisions and survey edges
in \S~\ref{incompleteness}.  The group catalogs are published in electronic tables
and their contents are described in \S~\ref{catalog}.  Finally, in \S~\ref{multiplicity},
we present our measured group multiplicity function.  We will use this to constrain
the HOD in future work.  We summarize our results in \S~\ref{summary}.

%#############################################################################

\section{Data} \label{data}

\subsection{SDSS}

The SDSS is a large imaging and spectroscopic survey that is mapping two-fifths of the 
Northern Galactic sky and a smaller area of the Southern Galactic sky, using a 
dedicated 2.5 meter telescope \citep{gunn_etal_06} at Apache Point, New Mexico.  
The survey uses a photometric camera \citep{gunn_etal_98} to scan the sky 
simultaneously in five photometric bandpasses \citep{fukugita_etal_96,smith_etal_02} 
down to a limiting $r$-band magnitude of $\sim22.5$.  The imaging data are processed 
by automatic software that does astrometry \citep{pier_etal_03}, source identification, 
deblending and photometry \citep{lupton_etal_01,lupton_05}, photometric calibration
\citep{hogg_etal_01,smith_etal_02,tucker_etal_05}, and data quality assessment
\citep{ivezic_etal_04}.  Algorithms are applied to select spectroscopic targets for
the main galaxy sample \citep{strauss_etal_02}, the luminous red galaxy sample
\citep{eisenstein_etal_01}, and the quasar sample \citep{richards_etal_02}.
The main galaxy sample is approximately complete down to an apparent $r$-band
Petrosian magnitude limit of $<17.77$.  Targets are assigned to spectroscopic plates
using an adaptive tiling algorithm \citep{blanton_etal_03a}.  Finally,
spectroscopic data reduction pipelines produce galaxy spectra and redshifts.

We use the large-scale structure sample \texttt{sample14} from the NYU Value 
Added Galaxy Catalog (NYU-VAGC; \citealt{blanton_etal_04a}) as our primary galaxy 
sample.  Galaxy magnitudes are corrected for Galactic extinction 
\citep{schlegel_etal_98} and absolute magnitudes are k-corrected 
\citep{blanton_etal_03b} and corrected for passive evolution \citep{blanton_etal_03c}
to rest-frame magnitudes at redshift $z=0.1$.  A significant fraction of 
the sample that we use was made publicly available with the SDSS Data Release~3 
\citep{abazajian_etal_05}.  

%-----------------------------------------------------------------------
\begin{table}
\begin{center}
\centerline{\small Table~1. Volume-limited Sample Parameters}
\begin{tabular}[t]{lccccc}
\tableline
\tableline
Name & $\zmin$ & $\zmax$ & $<\Mr$ & $\Ng$ & $\ng$ \\
\tableline
$Mr20$ & 0.015 & 0.100 & -19.9 & 57138 & 0.00673 \\
$Mr19$ & 0.015 & 0.068 & -19.0 & 37820 & 0.01396 \\
$Mr18$ & 0.015 & 0.045 & -18.0 & 18895 & 0.02434 \\
\tableline
\label{tab:samples}
\end{tabular}
\end{center}
Note---Absolute magnitude thresholds listed are for $\zmax$.  $\ng$ is in 
units of $\hden$.
\end{table}
%-----------------------------------------------------------------------

The galaxy redshift sample has an incompleteness due to the mechanical restriction 
that spectroscopic fibers cannot be placed closer to each other than their own 
thickness.  This fiber collision constraint makes it impossible to obtain redshifts 
for both galaxies in pairs that are closer than $55''$ on the sky.  In the case of a 
conflict, the target selection algorithm randomly chooses which galaxy gets a fiber
\citep{strauss_etal_02}.\footnote{In cases where a target galaxy fiber collides with a 
target quasar fiber, priority is always given to the quasar, but such collisions
only constitute $\sim 5\%$ of all cases.}  Spectroscopic plate overlaps alleviate this 
problem to some extent, but fiber collisions still account for a $\sim 6\%$ 
incompleteness in the main galaxy sample.  Since this incompleteness is most severe in 
regions of high galaxy density, it is necessary to correct for it in 
studies of groups and clusters.  We correct for fiber collisions by giving each 
collided galaxy the redshift of its nearest neighbor on the sky (usually the galaxy 
it collided with), and we show in \S~\ref{incompleteness} that this procedure is 
adequate for our purposes.
Putting collided galaxies at the redshifts of their nearest neighbors will
cause some nearby galaxies to be placed at high redshift, artificially making 
their estimated luminosities very high.  Since the abundance of highly luminous 
galaxies is low, this contamination can become a significant fraction of all highly 
luminous galaxies.  For this reason, we also give collided galaxies the magnitudes
(in addition to the redshifts) of their nearest neighbors.  The resulting luminosity
distribution is thus unbiased.

There is some additional incompleteness due to bright foreground stars
blocking background galaxies, but this is at the $\sim 1\%$ level.  In order
to limit the effects of incompleteness on our group identification, we
restrict our sample to regions of the sky where the completeness (ratio of
obtained redshifts to spectroscopic targets) is greater than $90\%$.  Our final
sample covers 3495.1 square degrees on the sky and contains 298729 galaxies.

\subsection{Volume-limited Samples}

In this and subsequent papers, we are primarily interested in using galaxy
groups to constrain the properties of galaxies as a function of their 
underlying dark matter halo mass.  It is therefore important that the population
of galaxies constituting the groups is homogeneous within the sample volume.
For this reason, we construct volume-limited subsamples of the full SDSS redshift 
sample that are each complete in a specified redshift range down to a limiting 
$r$-band absolute magnitude threshold.  We construct each sample by choosing 
redshift limits $\zmin$ and $\zmax$, and only keeping galaxies whose 
evolved, redshifted spectra would still make the redshift survey's apparent 
magnitude and surface brightness cuts at the limiting redshifts of the sample.
Since the apparent magnitude limit of the redshift sample varied across the sky in 
the commissioning phases of the survey, we cut the $r$-band magnitude limit from 
$\sim17.77$ back to 17.5.  This more conservative limit is uniform across the sky.

%-----------------------------------------------------------------------
\begin{figure}[t]
\epsscale{1.0}
\plotone{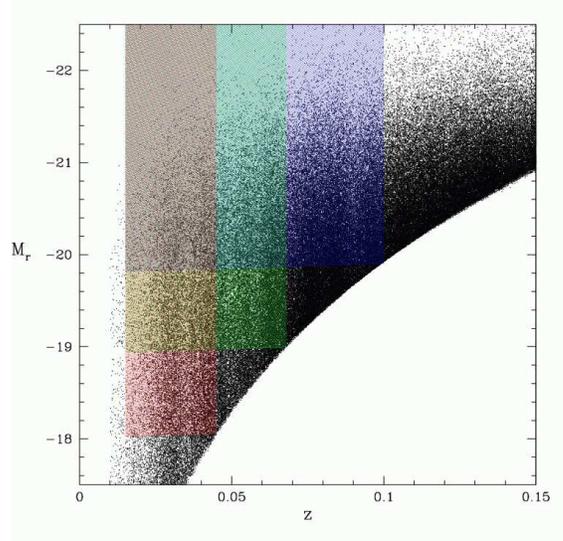}
\caption{
Absolute $r$-band magnitude vs. redshift for galaxies in the SDSS redshift survey,
highlighting the three volume-limited samples used for group identification. 
The three samples contain galaxies in the redshift ranges $0.015-0.1$, 
$0.015-0.068$, and $0.015-0.045$ and are complete for galaxies with r-band 
absolute magnitudes brighter than $-19.9$, $-19$, and $-18$, correspondingly.
The absolute magnitude threshold for a given volume-limited sample evolves 
with redshift in order to account for passive luminosity evolution of the 
galaxy population.
}
\label{fig:vollim}
\end{figure}
%-----------------------------------------------------------------------
%-----------------------------------------------------------------------
\begin{figure*}[t]
\includegraphics[scale=0.7,angle=-90]{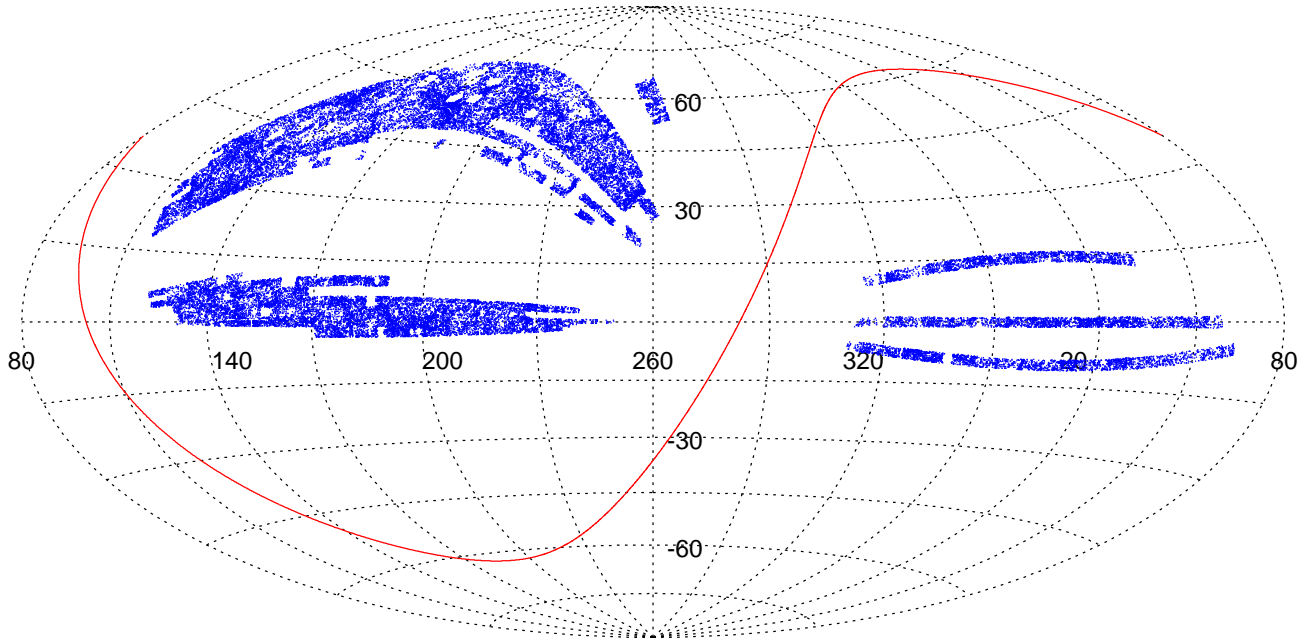}
\caption{
Hammer (equal area) projection (in equatorial coordinates) of the SDSS volume-limited 
sample that goes out to redshift $0.1$.  Points represent galaxies in the 
sample.  The solid curve shows the location of the Galactic plane.
}
\label{fig:skyvollim}
\end{figure*}
%-----------------------------------------------------------------------

We construct three such volume-limited samples.  Figure~\ref{fig:vollim} shows 
these samples in the luminosity-redshift plane.  Each dot in the figure shows
a galaxy in the SDSS redshift survey.  The sharp cutoff curve along the lower-right
part of the plot shows our $r=17.5$ apparent magnitude limit.
We select three redshift ranges for our volume-limited samples: $0.015-0.1$,
$0.015-0.068$, and $0.015-0.045$.  These samples are complete down to absolute
$r$-band magnitudes of $\Mr<-19.9$, $-19$, and $-18$, respectively.\footnote{All 
absolute magnitudes are quoted for $\Omegam=0.3$, $\Omegal=0.7$, and a value of the 
Hubble constant $h \equiv H_0 / 100\Hunits = 1$.  For other values of $H_0$, one 
should add $5\log h$ to the quoted absolute magnitudes.}  We refer to these samples 
as $Mr20$, $Mr19$, and $Mr18$, henceforth.  Regions of the plot that make it into 
these three samples are shown in blue, green, and red, respectively.  The limiting 
absolute magnitude of each sample changes slightly with redshift due to the passive
evolution corrections applied to galaxy luminosities: as a galaxy is moved to
the outer edge of a given volume-limited sample, its luminosity increases
somewhat, allowing lower redshift galaxies to make it into the sample at lower
luminosities than they do at higher redshifts.  We choose the first limiting
redshift of $\zmax=0.1$ because this yields the largest possible volume-limited
sample (largest number of galaxies).  We choose lower redshift samples in 
order to probe galaxy populations less luminous than $\Lstar$.  We use a lower
redshift limit of $0.015$ for all three samples to alleviate some of the problems
associated with obtaining accurate photometry of nearby highly extended galaxies.
The redshift limits, luminosity thresholds at $\zmax$, number of galaxies, and space 
densities of these samples are listed in Table~1. 

Figure~\ref{fig:skyvollim} shows a Hammer (equal area) projection (in equatorial 
coordinates) of sample $Mr20$.  Points represent galaxies in the sample.  The curve 
shows the location of the Galactic plane.  The figure illustrates the patchy and 
non-uniform nature of the sample footprint on the sky, which has irregular edges, 
as well as multiple holes.  This irregularity exacerbates systematic errors due to edge 
effects.  We deal with incompleteness due to edge effects in \S~\ref{incompleteness}.

Figure~\ref{fig:slicevollim} shows an equatorial slice through sample $Mr20$. The 
slice is $4^\circ$ thick and each point shows the RA and redshift of a galaxy in 
the sample.  Prominent in this projection of the data is the the giant supercluster 
at $z\sim0.08$ at the left end of the Sloan Great Wall of Galaxies, which extends from 
longitude 132 degrees (at $z\sim0.05$) to longitude 210 degrees (at $z\sim0.08$) (See 
\citealt{gott_etal_05}).

%#############################################################################

\section{Mock Catalogs} \label{mocks}

Our main scientific motivation for constructing group catalogs from the SDSS
data requires that identified groups most closely resemble systems of galaxies 
that occupy a common dark matter halo.  Moreover, it is important that we
statistically quantify the degree to which our groups do not satisfy this
criterion.  For both these reasons, it is imperative that we use mock galaxy
catalogs that are constructed by populating dark matter halos in N-body
simulations with mock galaxies.  The N-body simulations must satisfy two
basic conditions: they must contain a large enough volume to fit our largest
volume-limited sample, $Mr20$, and they must resolve the smallest mass halos 
that can host a galaxy in our least luminous volume-limited sample, $Mr18$.
HOD fits to the SDSS two-point correlation function of galaxies suggest that 
the minimum dark matter halo mass that can host a galaxy of luminosity
$\Mr\sim -18$ is approximately $2\times 10^{11}\hMsun$ 
\citep{zehavi_etal_05,tinker_etal_05}.  Requiring that a halo contain at least 
forty dark matter particles to be resolved means that we need N-body simulations 
with particle masses less than $5\times 10^{9}\hMsun$.

We use a series of N-body simulations of a $\Lambda$CDM cosmological model, 
with $\Omegam=0.3$, $\Omegal=0.7$, $\Omegab=0.04$, 
$h\equiv H_0/(100~\mathrm{km~s}^{-1}~\mathrm{Mpc}^{-1})=0.7$, $n_s=1.0$, and
$\sigma_8=0.9$.  This model is in good agreement with a wide variety of
cosmological observations (see, e.g., \citealt{spergel_etal_03,tegmark_etal_04b,
abazajian_etal_05b}).
Initial conditions were set up using the transfer function calculated 
for this cosmological model by CMBFAST \citep{seljak_zaldarriaga_96}.  The 
simulations were run at Los Alamos National Laboratory (LANL) using the 
Hashed-Oct-Tree (HOT) code \citep{warren_Salmon_93}.  We use a total of
six independent simulations of varying size and resolution, which we refer
to as \texttt{LANL1-6}.  The size of box $\Lbox$,
number of particles 
$N_\mathrm{p}$, and resulting particle mass $m_\mathrm{p}$ for each 
simulation are listed in Table~2.  The gravitational force
softening is $\epsilon_{\rm grav}=12\hkpc$ (Plummer equivalent).

%-----------------------------------------------------------------------
\begin{table*}
\begin{center}
\centerline{\small Table~2. Mock Catalog Parameters}
\begin{tabular}{l|cccc|cccc}
\tableline
\tableline
&
\multicolumn{4}{c|}{N-body} &
\multicolumn{4}{c}{HOD} \\
Mock & Name & $\Lbox$ & $N_\mathrm{p}$ & $m_\mathrm{p}$ & $\Mmin$ & $\Mcut$ & $M_1$ & $\alpha$ \\
 & & ($\hmpc$) & & ($10^9\hMsun$) & ($10^{11}\hMsun$) & ($10^{13}\hMsun$) & ($10^{12}\hMsun$) & \\
\tableline
\texttt{LANL1.Mr20} & \texttt{LANL1} & $384$ & $1024^3$ & 4.39 & 10.0 & --- & 25.0 & 1.1 \\
\texttt{LANL1.Mr20b} & & & & & 9.08 & 1.14 & 12.3 & 0.9 \\
\texttt{LANL1.Mr19} & & & & & 3.7 & --- & 8.2 & 1.0 \\
\texttt{LANL1.Mr18} & & & & & 1.9 & --- & 3.4 & 0.9 \\
\tableline
\texttt{LANL2.Mr20} & \texttt{LANL2} & $384$ & $1024^3$ & 4.39 & 10.0 & --- & 25.0 & 1.1 \\
\texttt{LANL2.Mr20b} & & & & & 9.08 & 1.14 & 12.3 & 0.9 \\
\texttt{LANL2.Mr19} & & & & & 3.7 & --- & 8.2 & 1.0 \\
\texttt{LANL2.Mr18} & & & & & 1.9 & --- & 3.4 & 0.9 \\
\tableline
\texttt{LANL3.Mr20} & \texttt{LANL3} & $384$ & $1024^3$ & 4.39 & 10.0 & --- & 25.0 & 1.1 \\
\texttt{LANL3.Mr20b} & & & & & 9.08 & 1.14 & 12.3 & 0.9 \\
\texttt{LANL3.Mr19} & & & & & 3.7 & --- & 8.2 & 1.0 \\
\texttt{LANL3.Mr18} & & & & & 1.9 & --- & 3.4 & 0.9 \\
\tableline
\texttt{LANL4.Mr20} & \texttt{LANL4} & $400$ & $1280^3$ & 2.54 & 10.0 & --- & 25.0 & 1.1 \\
\texttt{LANL4.Mr20b} & & & & & 9.08 & 1.14 & 12.3 & 0.9 \\
\texttt{LANL4.Mr19} & & & & & 3.7 & --- & 8.2 & 1.0 \\
\texttt{LANL4.Mr18} & & & & & 1.9 & --- & 3.4 & 0.9 \\
\tableline
\texttt{LANL5.Mr20} & \texttt{LANL5} & $543$ & $1024^3$ & 12.4 & 10.0 & --- & 25.0 & 1.1 \\
\texttt{LANL5.Mr20b} & & & & & 9.08 & 1.14 & 12.3 & 0.9 \\
\tableline
\texttt{LANL6.Mr20} & \texttt{LANL6} & $768$ & $1024^3$ & 35.1 & 10.0 & --- & 25.0 & 1.1 \\
\tableline
\label{tab:mocks}
\end{tabular}
\end{center}
\end{table*}
%-----------------------------------------------------------------------

We identify halos in the dark matter particle distributions using a 
friends-of-friends algorithm with a linking length equal to $0.2$ times the
mean interparticle separation.  We then populate these halos with galaxies
using a simple model for the HOD of galaxies more luminous than a luminosity
threshold.  Every halo with a mass $M$ greater than a minimum mass $\Mmin$ gets a 
central galaxy that is placed at the halo center of mass and is given the mean halo 
velocity.  A number of satellite galaxies is then drawn from a Poisson distribution 
with mean $\Nsat = ((M-\Mmin)/M_1)^\alpha$, for $M\geq\Mmin$.  These satellite 
galaxies are assigned the positions and velocities of randomly selected dark matter 
particles within the halo.  In order to construct mock catalogs for each of our 
three volume-limited samples $Mr20$, 
$Mr19$, and $Mr18$, we select sets of 
values for the parameters $\Mmin$, $M_1$, and $\alpha$ that yield 
the observed \citet{zehavi_etal_05} galaxy-galaxy correlation functions for 
these samples.  These HOD parameter values are similar to the best-fit values
given by \citet{zehavi_etal_05} (they are slightly different because the
model for $\Nsat$ was different in that paper).  We refer to these sets of
mock catalogs with the suffixes \texttt{.Mr20}, \texttt{.Mr19}, and \texttt{.Mr18}.
In addition to these mock catalogs, we construct a set of 
catalogs for the $Mr20$ sample using an alternative HOD model, where the mean 
number of satellites in a halo of mass $M$ is
$\Nsat = \mathrm{exp}[-\Mcut/(M-\Mmin)] (M/M_1)^\alpha$, for $M>\Mmin$
(also used by \citealt{tinker_etal_05}).  We fix the value of the slope $\alpha$ 
to 0.9, which is lower than that for the \texttt{.Mr20} mocks, and we choose 
values for the remaining HOD parameters that yield the observed 
\citet{zehavi_etal_05} correlation function of $\Mr<-20$ galaxies.  We refer
to these sets of mock catalogs with the suffix \texttt{.Mr20b}.  The values
for all mock HOD parameters are listed in Table~2.
We construct ten realizations of each mock catalog listed in Table~2
by using different random number generator seeds when we (a) draw a number of 
satellite galaxies for each halo from a Poisson distribution of mean $\Nsat$,
and (b) select random dark matter halo particles to give their positions and 
velocities to these satellite galaxies.  The dispersion among the ten 
realizations for one mock catalog therefore represents the scatter among
possible observed states for a given halo distribution and HOD model.

We now have a set of mock catalogs containing galaxies in real space and in
the cubical geometry of the N-body simulations.  We refer to these as our 
``real-space cube mocks''.  We create a redshift-space version of these catalogs
by assuming the distant observer approximation and aligning the line-of-sight
along one of the axes of the simulation cubes.  We use the mock galaxies'
peculiar velocities to move them along the line-of-sight into redshift space.
We refer to the resulting mock catalogs as our ``redshift-space cube mocks''.
We use these real-space and redshift-space cube mocks to determine optimal
parameters for our group-finding algorithm.  We summarize this determination in 
\S\ref{groupfinder} and discuss details in the Appendix.

For the purpose of studying the effects of SDSS incompleteness on our measured
groups, as well as for obtaining estimates of the uncertainty in our measured
group multiplicity function, we also require mock catalogs that have the same
geometry as our SDSS volume-limited samples.  The total volume of our largest
sample, $Mr20$, is approximately $210^3\hvol$, which is more than six times 
smaller than any of our mock cubes.  However, the SDSS geometry is highly
irregular (as seen in Fig.~\ref{fig:skyvollim}) and can only be fully embedded in
a cube of much larger volume than the survey itself.  The $Mr20$ sample, for
example, has a maximum extent of $\sim600\hmpc$ when both the North and 
South Galactic portions are included.  In order to carve this sample geometry
out of our mock catalogs, we create mock cubes with eight times larger volume by
tiling each mock cube $2\times2\times2$.  Since the N-body simulations used
to construct the mocks were run with periodic boundary conditions, we can tile
the cubes without having density discontinuities at the boundaries.  We set the 
center of this tiled cube to be the origin and put galaxies into redshift space 
using the line-of-sight component of their peculiar velocities.  We then
compute RA, DEC, and redshift coordinates for every mock galaxy in the tiled 
cube.  Finally, we only keep galaxies whose coordinates on the sky would place 
them in regions of the SDSS survey that have completeness greater than $90\%$, 
and whose redshifts lie within the redshift limits of the specific volume-limited 
sample we are constructing mock catalogs for.

Since the volume of each simulation cube is at least six times larger than
our largest volume-limited sample $Mr20$, we try to carve out as many independent 
volumes with the $Mr20$ geometry as possible without too much overlap.  We do 
this by performing many sets of three rotations (one around each Cartesian axis) 
and testing how much overlap the resulting catalogs have with each other (i.e., 
how many common mock galaxies do they share). With the right combination 
of rotation angles, we can carve out two $Mr20$ mock catalogs that share fewer than 
$3\%$ of their galaxies with each other, but we cannot obtain more without 
significant overlap.  We create two such independent mock 
catalogs, with the correct SDSS geometry, from every one of the ten HOD 
realizations of the mock cubes listed in Table~2, except for the 
\texttt{LANL6.Mr20} mock.  This procedure yields 200 mock catalogs for 
the $Mr20$ sample (5 N-body simulations $\times$ 2 HOD models $\times$ 10 HOD 
realizations $\times$ 2 mocks per simulation cube), and 80 mock catalogs each 
for the $Mr19$ and $Mr18$ samples (4 N-body simulations $\times$ 1 HOD model 
$\times$ 10 HOD realizations $\times$ 2 mocks per simulation cube).

The final step in creating mock SDSS catalogs is to incorporate the fiber 
collision constraint.  We use a friends-of-friends algorithm to identify groups 
of mock galaxies that are linked together by the $55''$ minimum angular
separation of fibers.  We then select ``collided'' mock galaxies (whose redshifts 
will be unknown) in each such collision group in a way that minimizes the number of 
such galaxies.  For example, if a collision group contains three galaxies in a row, 
where the first is closer than $55''$ from the second and the second is closer than 
$55''$ from the third, but the first is more than $55''$ from the third, we will 
always select the middle galaxy to be the collided one.  In cases where multiple 
choices yield the same number of collided galaxies, we select randomly (e.g., in 
collision groups with only two galaxies).  This procedure is designed to mimic
the tiling code that assigns spectroscopic fibers to SDSS target galaxies
\citep{blanton_etal_03a}.  If we perform this operation on the \texttt{.Mr20} catalogs
we end up with only $\sim 3\%$ of mock galaxies being tagged as collided.  This
is about half the fraction of SDSS galaxies in our $Mr20$ sample that don't have 
measured redshifts due to fiber collisions.  The reason for this discrepancy is 
that galaxies in the $Mr20$ volume-limited sample do not only collide with each 
other; they also collide with galaxies more luminous than $\Mr\sim-20$ at redshifts 
higher than the sample limit $z=0.1$ and galaxies less luminous than  $\Mr\sim-20$ at 
lower redshifts.  Most of these additional galaxies that can collide with a given
galaxy in $Mr20$ are uncorrelated background or foreground galaxies.  It is 
therefore sufficient to model them as a background screen of galaxies on the sky
that have an angular correlation function equal to the mean for all SDSS galaxies.
For this purpose, we use the very large volume \texttt{LANL6.Mr20} cube mock.
We use \texttt{LANL6.Mr20} to construct a ``screen'' catalog with the correct SDSS 
angular geometry and a variable outer redshift limit, and superpose it onto each 
of our \texttt{.Mr20}, \texttt{.Mr19}, and \texttt{.Mr18} mock catalogs.  We then
allow all galaxies to collide with each other and keep track of collided mock galaxies.
We set the outer redshift limit of the screen catalog to the value that results in 
$\sim 6\%$ of mock galaxies being tagged as collided.  We find that we need 
approximately seven times more galaxies in the screen catalog than in the mocks
in order to achieve this collided fraction.  

Using this approach we construct three versions of every mock catalog described
above: a version with no fiber collisions applied (``true'' version), a version 
where collided galaxies have no redshifts and are dropped out of the mock catalog
altogether (``uncorrected'' version), and a version where collided galaxies
are assigned the redshift of the galaxy they collided with (``corrected'' version).
These mock catalogs allow us to test the effects of fiber collisions on our 
measured group multiplicity function (discussed in \S~\ref{incompleteness}.)

%#############################################################################

\section{Group-Finding Algorithm} \label{groupfinder}

We wish to identify galaxy groups primarily in order to measure the group
multiplicity function and use it to constrain the HOD of galaxies as a function 
of galaxy properties.
This goal places a number of demands on the group-finding algorithm:
(1) It should identify galaxy systems that occupy the same dark matter halos with 
the least possible merging of different halos into the same group and the least 
possible splitting of individual halos into multiple groups. (2) It should
produce a group multiplicity function that is unbiased with respect to the
halo multiplicity function. (3) It should be simple and well-defined so that the
statistical and systematic uncertainty in the measured group multiplicity function
can be accurately characterized.  (4) It should use only the spatial positions
of galaxies in redshift space to identify groups, and not galaxy properties
such as color or luminosity.  These requirements point to an algorithm that
uniquely identifies density enhancements in redshift space.

We adopt the simple and well understood friends-of-friends (FoF) algorithm, where
galaxies are recursively linked to other galaxies within a specified linking volume
around each galaxy.  The FoF algorithm has several attractive features.  First, for a 
given linking volume (usually specified by one linking length in real space and two 
linking lengths in redshift space), FoF produces a unique group catalog.  Second, it 
does not assume or enforce any particular geometry for groups (e.g., spherical), but 
rather identifies structures that are approximately enclosed by an isodensity surface
whose density is monotonically related to the linking lengths.  Third, the algorithm
satisfies a nesting condition: all the members of a group identified with one set of 
linking lengths are also members of the same group identified using larger linking
lengths.

The FoF algorithm has been used extensively to identify dark matter halos in N-body
simulations (e.g., \citealt{davis_etal_85}) and has been shown to produce halo
catalogs with mass functions that are close to universal (within $\sim 20\%$) for a wide 
range of epochs and cosmological models \citep{jenkins_etal_01}.  FoF has also been the
most used algorithm for identifying galaxy groups in redshift surveys 
\citep{huchra_geller_82,geller_huchra_83,nolthenius_white_87,ramella_etal_89,
moore_etal_93,ramella_etal_97,ramella_etal_99,tucker_etal_00,giuricin_etal_00,
ramella_etal_02,merchan_zandivarez_02,eke_etal_04a}, though alternative methods have 
also been used (see e.g., \citealt{tully_87,marinoni_etal_02,gerke_etal_05,
yang_etal_05}).  These FoF studies all used the same basic algorithm, but differed in 
their choices for linking lengths and in their methods for dealing with the varying 
density of galaxies inherent in flux-limited surveys.

We use the basic \citet{huchra_geller_82} algorithm, where two galaxies are linked to
each other if both their transverse and line-of-sight separations are smaller
than a given pair of projected and line-of-sight linking lengths, respectively.
Specifically, two galaxies $i$ and $j$ with angular separation $\theta_{ij}$ and
redshifts $z_i$ and $z_j$, have a projected separation $D_{\perp,ij}$ and a 
line-of-sight separation $D_{\parallel,ij}$ (both in $\hmpc$) given by 
\footnote{We use these simple equations, rather than the exact formulae for the
redshift-distance and angular diameter-distance relations because, at $z=0.1$ (the
outer limit of our sample), the difference between these formulae is less than $1\%$.}
\begin{eqnarray}
D_{\perp,ij} & = & (c/H_0)(z_i+z_j)~\mathrm{sin}(\theta_{ij}/2), \\
D_{\parallel,ij} & = & (c/H_0)|z_i-z_j|.
\end{eqnarray}
The two galaxies are then linked to each other if
\begin{equation}
D_{\perp,ij} \leq \bperp~\ng^{-1/3}
\end{equation}
and
\begin{equation}
D_{\parallel,ij} \leq \bpar~\ng^{-1/3},
\end{equation}
where $\ng$ is the mean number density of galaxies, and $\bperp$ and $\bpar$ are
the projected and line-of-sight linking lengths in units of the mean intergalaxy
separation.  Since we use volume-limited samples of SDSS galaxies, $\ng$ is
constant throughout the sample volumes, and thus the linking lengths are also
constant.

The resulting linking volume around each galaxy is very similar to a cylinder,
oriented along the line-of-sight, whose radius is equal to the projected linking
length and whose height is equal to twice the line-of-sight linking length.  It 
is not a perfect cylinder because its radius increases with redshift, making it 
slightly wider at the far end than at the near end, and its bases are slightly 
curved.  However, for the small linking lengths considered here, a cylinder is
a good approximation.  The FoF algorithm works recursively, whereby a galaxy is
linked to all its ``friends'', which are in turn linked to their ``friends'', 
etc., to yield a unique group of galaxies.

\subsection{Choice of Linking Lengths}

The most important ingredient of our group-finding algorithm is our choice
for the linking lengths $\bperp$ and $\bpar$.  If the linking lengths are
too small, then the group-finder will break up single halos into multiple
groups.  If the linking lengths are too large, then different halos will be
fused together into single groups.  There are no values for the linking lengths
that will work perfectly for every halo, even in real space.  In redshift space
this problem becomes substantially worse, since redshift-space distortions both
move halos and elongate them along the line-of-sight, often causing them to
overlap with each other.  The right choice of linking lengths depends on the 
purpose for which groups are being identified.  If we require a group catalog that 
is highly inclusive and groups together every galaxy inhabiting the same halo, then 
we will use larger linking lengths than if our goal is to minimize contamination by 
galaxies that come from different halos.  For our purposes, we wish to obtain a 
balance between being inclusive and reducing contamination, while producing groups 
that have an unbiased multiplicity function.

In order to find the right combination of linking lengths, we use the mock galaxy
catalogs described in \S~\ref{mocks}.  Specifically, we use the real-
and redshift-space cube mocks, which are constructed by applying simple HOD
models to the \texttt{LANL1} and \texttt{LANL4} N-body simulations.  Since we
know which mock galaxies occupy the same dark matter halos, we can evaluate
how well a particular choice of linking lengths recovers features of the halo
population.  The mocks that we use here have a cubical geometry, and we assume the
distant observer approximation when we put mock galaxies into redshift space.
We use the full cubical mocks rather than those with the correct SDSS geometry
because the full mocks have a much larger volume and thus better statistics.
Moreover, our goal is to find the best linking lengths for any redshift survey, and
we will deal with systematic effects specific to our SDSS sample geometry separately.
The FoF algorithm that we use is therefore slightly different from the one outlined
above, in that the linking volume is a perfect cylinder (i.e., $D_{\perp,ij}$ is
simply the projected distance between two mock galaxies).  

We run the FoF group-finder on the mock catalogs for a grid of linking length
values, and we study the properties of the resulting group catalogs.  Specifically,
we investigate four features of the recovered group distribution: (1) the group
multiplicity function compared to the ``true'' halo multiplicity function;
(2) The relation between the number of galaxies in a halo $\Ntrue$ and the
number of galaxies in its associated group $\Nobs$; (3) The distribution of projected
group sizes as a function of group richness compared to the ``true'' distribution
of projected halo sizes as a function of halo multiplicity; (4) The distribution of 
group velocity dispersions as a function of group richness compared to the ``true'' 
distribution of halo velocity dispersions as a function of halo multiplicity.

We check how each set of linking lengths performs in the above four tests, for
each of the four HOD model mock cubes (\texttt{.Mr20, .Mr20b, .Mr19, .Mr18}).
In the case of each HOD model, we average results over the 10 HOD realizations 
described in \S~\ref{mocks} and over the \texttt{LANL1} and \texttt{LANL4} N-body 
simulations.  We do this procedure for groups that are identified in both real space 
(for which there is only one linking length), and redshift space.  These tests are
described in detail in the Appendix.  Here we summarize the main results.

In real space, a linking length choice of $b=0.2$ yields galaxy groups with ten or 
more members that pass all four tests listed above.  Groups with $N<10$ show
systematic deviations in abundance, multiplicity, projected sizes, and velocity
dispersions from the corresponding halos with $N<10$.  The choice of $b=0.2$ is not 
surprising, given that the same linking length was used to identify halos in the
N-body simulations.  It is also not surprising that the group-finding fails the
tests for small groups, where adding or losing a couple of galaxies makes a large
fractional difference to the group size.  The threshold of $N\sim 10$ is independent
of the underlying dark matter halo mass.  This means that we can push the regime in
which the groups are reliable to lower mass systems by using a lower luminosity
sample (where each halo will contain more galaxies).  Of course, the change of 
luminosity threshold comes at the expense of statistical power, since low luminosity 
samples have smaller volumes than high luminosity samples.  The number of groups in a 
volume-limited sample scales roughly with the number of galaxies, and a luminosity 
threshold near the characteristic luminosity $L_*$ maximizes this number.

In redshift space the situation is more complicated.  No set of transverse and
line-of-sight linking lengths is able to produce groups that pass all four tests
listed above, even for large size groups.  Figure~\ref{fig:linkinglengths.hod20}
summarizes our tests for the \texttt{.Mr20} HOD model mocks.  Results for the
other HOD models are similar and are shown in the Appendix.  The figure shows
regions (shaded) of the two-dimensional linking length space ($\bpar$ vs. $\bperp$) 
that pass each of our four tests.

\subsubsection{Multiplicity Function}

\begin{figure}[t]
\epsscale{1.0}
\plotone{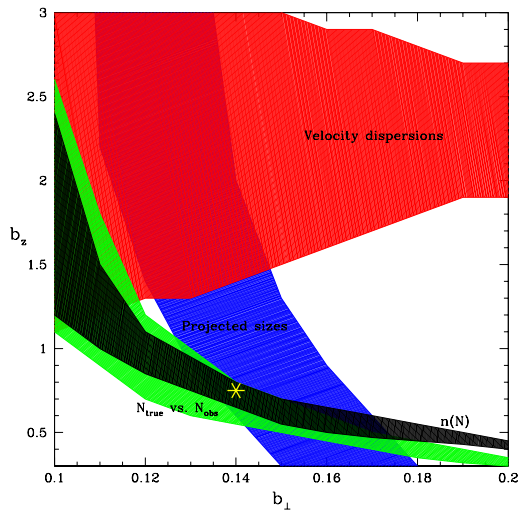}
\caption{
Regions of the FoF linking length parameter space that do well in recovering galaxy 
groups that have similar properties to their parent halos.  Each shaded region
shows the combination of perpendicular and line-of-sight FoF linking lengths that
are successful in recovering a particular feature of the group distribution, measured 
using mock galaxy catalogs.
The four features are: (a) the group multiplicity function ({\it black region}); (b)
the relation between halo and group richness for halos and groups that are matched 
one-to-one ({\it green region}); (c) the projected sizes of groups as a function of 
group richness ({\it blue region}); (d) the line-of-sight velocity dispersion of
groups as a function of group richness ({\it red region}).
The yellow star denotes the FoF parameters that we apply to identify groups in the
SDSS.  
}
\label{fig:linkinglengths.hod20}
\end{figure}

The dark and thin shaded region in Figure~\ref{fig:linkinglengths.hod20}, labeled 
$n(N)$, shows linking lengths that pass the group multiplicity function test.  In 
other words, these linking lengths yield mock group catalogs whose multiplicity 
functions are unbiased relative to the ``true'' input halo multiplicity function, 
in the regime $N\geq 10$.  In this case, ``unbiased'' means that the shape of the 
multiplicity function is on average the same as the ``true'' shape and its amplitude 
is within $10\%$ of the ``true'' amplitude.  Linking length values that lie along 
the upper boundary of the shaded region (e.g, the values $\bperp=0.11$, $\bpar=1.5$) 
yield multiplicity functions that are $10\%$ too high in amplitude, whereas values 
that lie along the lower boundary yield multiplicity functions whose amplitudes are 
$10\%$ too low.  These results show that an increase in either linking length
generally leads to an increase in the multiplicity function for $N\geq 10$.  This
increase is compensated for by a corresponding decrease in the abundance of isolated
(i.e., $N=1$) and low $N$ groups.  The shaded region appears to be close to horizontal 
only because the vertical axis is highly compressed with respect to the horizontal 
axis.

%-----------------------------------------------------------------------
%-----------------------------------------------------------------------

\subsubsection{$\Ntrue$ vs. $\Nobs$}

The group multiplicity function is an average statistic showing the abundance 
of all groups as a function of $N$.  It is therefore possible, in principle, for it 
to be unbiased relative to the halo multiplicity function, without the relation 
between individual halo multiplicities and their recovered group multiplicities
being correct.  For this reason, we also require that the group-finder yield an
unbiased relation between the multiplicity of individual halos, $\Ntrue$, and their 
recovered groups, $\Nobs$.  In order to check this, we must match input halos to 
recovered groups in a one-to-one way.  There are many ways to do this matching, and no 
one way is more correct than another.  For example, a halo can be associated with the 
group that contains most of its galaxies, or the group that contains its central 
galaxy, or the group whose centroid is closest to the halo center.  We associate 
each halo to the group that contains its central galaxy.  When two or more halos are
matched to the same group, we choose the halo that shares the largest number of 
common galaxies with the group.  Halos that are not associated with any group are
considered ``undetected,'' and groups that are not associated with any halo (because
they don't contain any halo central galaxies) are considered ``spurious''.

The light (and green) shaded region in Figure~\ref{fig:linkinglengths.hod20} that 
roughly tracks and is slightly wider than the $n(N)$ region shows linking lengths that 
pass the $\Ntrue$ vs. $\Nobs$ test.  In other words, these linking lengths yield mock 
group catalogs with an unbiased median relation between $\Ntrue$ and $\Nobs$ for 
associated halos and groups, in the regime $N\geq 10$.  We consider the relation to 
be unbiased if its slope is within $10\%$ of unity.  Linking length values that lie 
along the upper boundary of the shaded region yield associated halos and groups with 
a median relation $\Ntrue=1.1\Nobs$, whereas values that lie along the lower boundary 
yield the relation $\Ntrue=0.9\Nobs$.  As expected, most linking lengths that pass
the multiplicity function test also pass the $\Ntrue$ vs. $\Nobs$ test.  This breaks
down, however, for values of $\bperp$ greater than 0.16-0.17.

\subsubsection{Projected Sizes}

The (blue) shaded region in Figure~\ref{fig:linkinglengths.hod20}, labeled 
``Projected sizes'', shows linking lengths that pass the projected sizes test.  These 
linking lengths yield mock groups with an unbiased median relation between rms 
projected size and group multiplicity $N$, in the regime $N\geq 10$.  We consider the 
relation to be unbiased if it is within $10\%$ of the ``true'' relation between
median rms projected halo size and halo multiplicity.  This shaded region is roughly
vertically oriented because the projected linking length $\bperp$ affects the 
projected sizes of groups much more than the line-of-sight linking length $\bpar$.
Clearly, increasing $\bperp$ leads to galaxy groups with larger projected sizes.
The shaded region is not completely vertical, however, because increasing $\bpar$ 
also leads to larger projected size groups, albeit in a much less sensitive way.

\subsubsection{Velocity Dispersions}

The (red) shaded region in Figure~\ref{fig:linkinglengths.hod20}, labeled 
``Velocity dispersions'', shows linking lengths that pass the velocity dispersion
test.  These linking lengths yield mock groups with an unbiased median relation 
between group velocity dispersion and group multiplicity $N$, in the regime $N\geq 10$. 
We consider the relation to be unbiased if it is within $10\%$ of the ``true'' 
relation between median halo velocity dispersion and halo multiplicity.  This shaded 
region is roughly horizontally oriented because the line-of-sight linking length 
$\bpar$ affects the velocity dispersions of groups much more than $\bperp$.
Clearly, increasing $\bpar$ leads to galaxy groups with larger velocity dispersions.
The shaded region is not completely horizontal, because changing $\bperp$ also
affects the velocity dispersions of groups, though not consistently in the same
sense.

\subsubsection{Our Adopted Linking Lengths}

It is obvious from Figure~\ref{fig:linkinglengths.hod20} that no combination of
FoF linking lengths passes all four tests listed above.  We can choose linking
lengths that successfully recover the abundance and projected sizes, or the abundance 
and velocity dispersions of groups as a function of multiplicity, but not all three
simultaneously.  We can also choose linking lengths that successfully recover
both the projected sizes and velocity dispersions of groups as a function of
multiplicity, but since the multiplicity function of such groups is incorrect, the 
overall size and velocity dispersion distributions will also be incorrect.
This failure to recover all features of groups in redshift space is a fundamental 
shortcoming of the FoF group-finder when applied to redshift space.  Given that 
most redshift-space group-finding algorithms operate on very similar principles, 
i.e., they identify overdense regions that are elongated along the line-of-sight, 
it is likely that this shortcoming is shared by other group-finders as well.  To 
our knowledge, no group-finder has been shown to pass all four of the tests 
considered here for a single choice of parameters.

Figure~\ref{fig:linkinglengths.hod20} shows that in order to recover groups with 
unbiased velocity dispersions, the line-of-sight linking length must be substantially
larger than the mean intergalaxy separation.  With $\bpar$ that large, groups are
bound to be linked together along the line-of-sight.  
The only way to then obtain groups with the correct multiplicity function is to have 
a transverse linking length small enough that galaxies in the outer parts of halos are 
not included in the recovered groups.  The resulting groups bear little physical 
resemblance to their parent halos.  If, on the other hand, we recover groups with 
unbiased projected sizes, then the groups will be missing some of their fastest moving 
galaxies and this decrease in multiplicity will be compensated by including as group 
members a few galaxies in the infall regions of halos.  These groups are much more 
physically similar to their parent halos.  For this reason, we choose to sacrifice 
velocity dispersions, rather than projected sizes, when selecting values for the FoF 
linking lengths.   

Figure~\ref{fig:linkinglengths.hod20} shows the linking length values that we
adopt and use in this paper (yellow star).  These values are
\begin{equation}
\bperp=0.14, \qquad \bpar=0.75 ~.
\end{equation}
Our mock catalog tests show that the FoF algorithm with these linking lengths finds
galaxy groups with $N\geq 10$ that have: (1) an unbiased multiplicity function; 
(2) an unbiased median relation between the multiplicities of groups and their 
associated halos; (3) a spurious group fraction of less than $\sim 1\%$; (4) a halo 
completeness (fraction of halos that are associated one-to-one with groups) of more 
than $\sim 97\%$; (5) the correct projected size distribution as a function of 
multiplicity; (6) a velocity dispersion distribution that is $\sim 20\%$ too low
at all multiplicities.  These results hold for all of the mock catalogs that we 
have used (see results for other HOD models in the Appendix) and are thus not very 
sensitive to the HOD model assumed or to the specific realization of the underlying 
density field.  We note that our adopted group-finder only has the above properties
when dark matter halos are defined using a FoF algorithm with a linking length of
0.2 times the mean interparticle separation, since that was the definition used to 
construct our mock catalogs.  A different halo definition (such as FoF using a 
different linking length, or a spherical overdensity halo-finder) will result in a 
different optimal group-finder.

Previous FoF group analyses have used different linking lengths.  For example, 
\citet{eke_etal_04a} adopt $\bperp=0.13, \bpar=1.43$ in their analysis of groups in the 
2dF Galaxy Redshift Survey (2dFGRS; \citealt{colless_etal_01}).  With a similar 
transverse linking length but much larger line-of-sight linking length than used here, 
this parameter combination yields unbiased projected sizes and velocity dispersions, 
but it overpredicts the abundances of halos by $20-30\%$ at large multiplicities (see 
Figure~\ref{fig:linkinglengths.hod20}).  These groups are thus poorly suited to our 
primary objective of using group abundances as a cosmological test.
\citet{yang_etal_05} and \citet{weinmann_etal_05} use a group-finder that assumes
a mass, radius, and velocity dispersion for each preliminary group and then includes or 
discards galaxies from the group based on these assumed properties (similar to a
matched filter technique).  This method might, in principle, be able to simultaneously
recover groups with unbiased abundances, projected sizes, and velocity dispersions -
at the expense of model independence - but this remains to be tested.

%#############################################################################

\section{Incompleteness} \label{incompleteness}

There are two main sources of incompleteness that will affect the richnesses of
groups, and hence the multiplicity function, in our SDSS group catalogs: fiber
collisions and survey edges.  Both these effects will prevent galaxies from
being included in some groups, and thus cause the richness of these groups to be 
underestimated.  These sources of incompleteness and their effects on the measured 
group multiplicity function must be accounted for.

\subsection{Fiber Collisions} \label{fibcols}

%-----------------------------------------------------------------------
\begin{figure}[t]
\epsscale{1.0}
\plotone{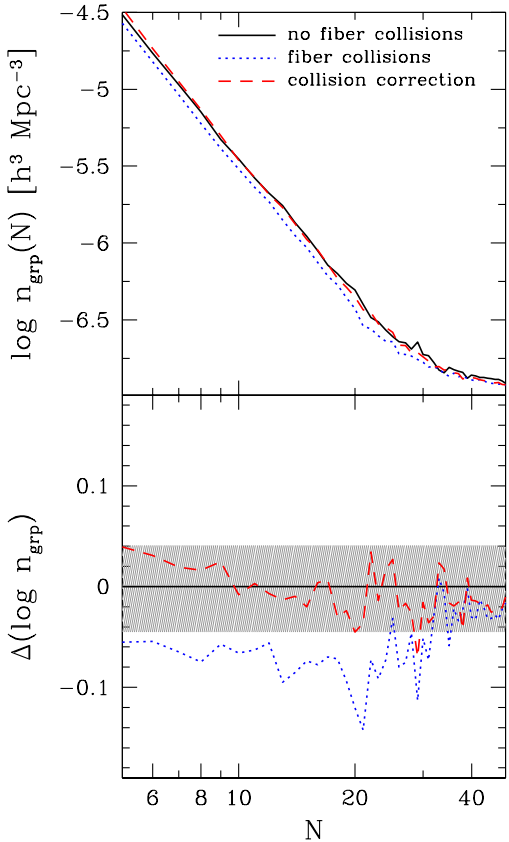}
\caption{
Effect of fiber collisions on the group multiplicity function measured using
mock SDSS galaxy catalogs, which are described in \S~\ref{mocks}.
The top panel shows the differential group multiplicity function for mock
catalogs that contain no fiber collisions and thus represent the ``true''
case ({\it solid black curve}), that lose galaxies due to fiber collisions as
in the SDSS survey ({\it dotted blue curve}), and that are corrected for fiber
collisions as described in \S~\ref{incompleteness} ({\it dashed red curve}).
The bottom panel shows the ratio of each case to the ``true'' one.  The shaded
region encloses $\pm 10\%$ deviations from the ``true'' multiplicity function.
These results are averaged over all of our \texttt{.Mr20} mock catalogs.
}
\label{fig:nbodyfibcol}
\end{figure}
%-----------------------------------------------------------------------

Fiber collisions cause an incompleteness that grows with the surface density of 
galaxies and is thus especially important in group and cluster studies.  Moreover,
the surface density in groups is likely a function of group richness.  The mean
surface density of a group of richness $N$, mass $M$, and radius $R$ scales like 
$\Sigma \sim N/R^2 \sim N/M^{2/3}$.  For a power-law relation between mean richness
and halo mass $N\sim M^\alpha$, the surface density is $\Sigma \sim N^{1-2/3\alpha}$.
This scaling relation is clearly a crude approximation, but it illustrates that
the incompleteness due to fiber collisions likely varies with group richness and 
can thus affect both the amplitude and slope of the multiplicity function.

We use the 100 \texttt{LANL1-5.Mr20} mock catalogs (5 N-body simulations $\times$ 
10 HOD realizations $\times$ 2 mocks per simulation cube) to assess the impact of 
fiber collisions on the group multiplicity function.   We apply the group-finder 
described in \S~\ref{groupfinder} to the ``uncorrected'' and ``true'' versions of 
these mock catalogs and measure the resulting multiplicity functions. 
Figure~\ref{fig:nbodyfibcol} shows these multiplicity functions averaged over all 
the mock catalogs.  The figure shows that dropping collided galaxies from
the sample lowers the amplitude of the multiplicity function by more than 10\% and 
also slightly changes its slope.  The amplitude drops because some groups in each
richness bin lose galaxies and are thus shifted to lower $N$ bins.  There are also
some groups from higher $N$ bins that are shifted into these bins, but their number
is smaller than the number of groups lost because the abundance of groups drops
steeply with increasing $N$.

\citet{zehavi_etal_05} show that the effect of fiber collisions on the galaxy
two-point correlation function can be successfully corrected for by including 
each collided galaxy at the redshift of its nearest neighbor.  We apply the same 
correction to our mock catalogs to produce a set of ``corrected'' mocks. 
Figure~\ref{fig:nbodyfibcol} shows that this correction works very well in the regime
$N\geq 10$, and we therefore adopt it for our group identification.

%-----------------------------------------------------------------------
\begin{figure*}[t]
\epsscale{0.8}
\plotone{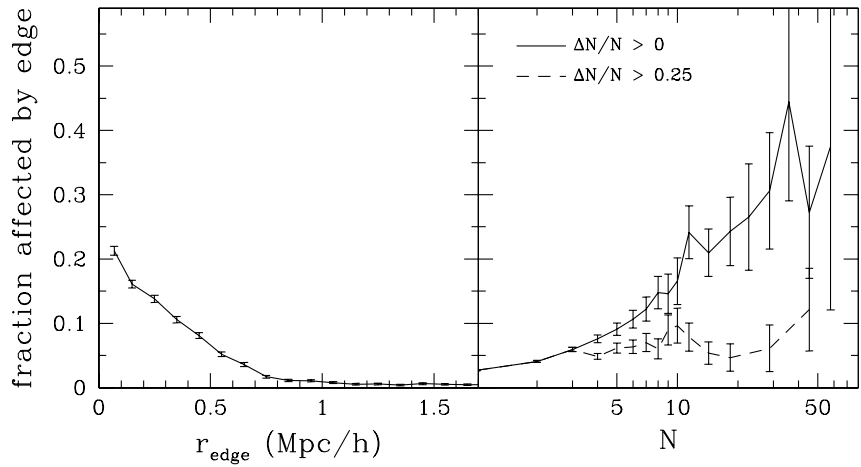}
\caption{
Fraction of groups affected by survey edges, measured using mock SDSS galaxy 
catalogs (described in \S~\ref{mocks}).  Groups are considered affected by
edges if they lose any galaxies that would have been included in the absence of
edges.  The panels show the edge fraction of groups in bins of the distance from
their centroids to the closest edge $\redge$ ({\it left panel}) and group richness 
$N$ ({\it right panel}).  The right panel also shows the fraction of groups
that lose more than $25\%$ of their member galaxies due to edges ({\it dashed
curve}).  These results are averaged over four independent \texttt{.Mr20} mock 
catalogs.
}
\label{fig:nbodyedgestats}
\end{figure*}
%-----------------------------------------------------------------------

\subsection{Survey Edges} \label{edges}

Groups that are identified near the edges of a given sample could be missing
galaxies that are located just outside the sample.  Similar to fiber collisions,
edge effects always shift groups from higher to lower richness.  Moreover, large and 
extended groups have a higher probability of being affected by edges than do small 
and compact groups because they can straddle an edge while being further away from it.
Edge effects are most severe when the ratio of a sample's surface area to its enclosed
volume is high.  Figure~\ref{fig:skyvollim} shows that the SDSS sample has a highly 
irregular footprint on the sky, which implies a high surface-to-volume ratio.  Edge
effects are, therefore, potentially severe in our samples.  When the SDSS survey is 
complete and the gap in the North Galactic cap is filled in, edge effects will be
much less important.

We can measure the effects of edges using our mock catalogs, since we know what 
galaxies lie on the other side of edges.  For every group identified in our 
\texttt{LANL1-5.Mr20} mock catalogs, we determine how many galaxies are missing due to 
edges.  An edge can lie either in the perpendicular direction, or along the 
line-of-sight due to a sample's redshift limits.

\begin{figure}[t]
\epsscale{1.0}
\plotone{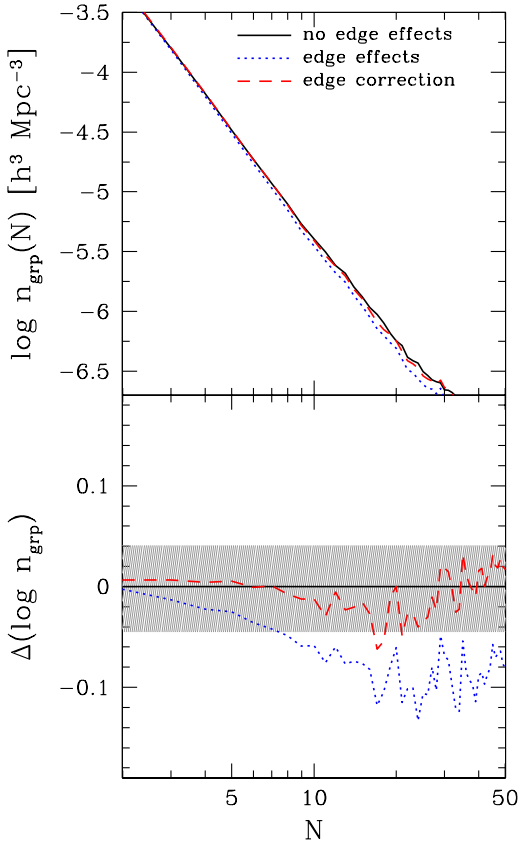}
\caption{
Effect of survey edges on the group multiplicity function measured using
mock SDSS galaxy catalogs (described in \S~\ref{mocks}).  The figure shows the
group multiplicity function for mock catalogs that contain no edge effects and
thus represent the ``true'' case ({\it solid black curve}), that contain edge effects
as in the SDSS survey ({\it dotted blue curve}), and that are corrected for edge
effects as described in \S~\ref{incompleteness} ({\it dashed red curve}).  All other
features as in Fig.~\ref{fig:nbodyfibcol}.  These results are averaged over all of our
\texttt{.Mr20} mock catalogs.
}
\label{fig:nbodyedge}
\end{figure}

The solid curve in the right panel of Figure~\ref{fig:nbodyedgestats} shows the fraction
of mock groups that are missing one or more galaxies due to edges, as a function of
group richness $N$.  The affected fraction climbs from 10\% to 40\% as $N$ goes from
5 to 50.  Edges clearly affect a large fraction of high richness groups in our 
sample, but counting a group as affected if it loses only a single galaxy is a very
conservative test.  It makes more sense to calculate the fraction of groups that lose 
a fixed fraction of their galaxies, rather than just a single galaxy.  The dashed curve
in the same panel shows the fraction of groups that lose 25\% or more of their 
galaxies.  The affected fraction defined this way is $\sim10\%$, roughly independent
of richness.  Figure~\ref{fig:nbodyedge} shows the effect of edges on the multiplicity 
function (blue curve).  The effect of edges on the abundance of mock groups grows from 
zero at $N=2$ to approximately 20\% at $N=50$.  It is, therefore, very important to 
correct for edges, since they systematically change the shape of the multiplicity 
function and, hence, the derived HOD.

We measure the shortest distance of every galaxy from the survey edges by laying down
points around each galaxy at successively larger radii and checking if they also lie
within our sample volume.  The smallest radius at which points fall outside the sample
volume is the distance of the galaxy from the edge.  Any group that contains at least 
one galaxy within a linking length from the edge, whether it is a projected 
linking length in the tangential direction or a line-of-sight linking length in the 
redshift direction, is potentially affected, since there could be galaxies on the other 
side that would be linked to the same group.  One possible way to deal with edges is
to throw out all such groups.  This is a very conservative solution, since it ensures 
that all groups in our final sample are uncontaminated by edges.  However, it is tricky 
to estimate the new effective volume of the sample, which is necessary for measuring the 
multiplicity function.  Moreover, the effective volume for large groups will be smaller 
than that for small groups.  Another possibility is to keep all groups, but somehow 
correct the multiplicities of those that are potentially affected by edges.  This 
solution has the advantage that no groups are lost, but it is once again difficult to 
estimate the effective volume of the sample, even if all multiplicity corrections are 
exactly right.  A third possibility is to reject all groups whose centers 
lie less than a minimum distance from the edge.  This correction has the advantage that 
it produces an unbiased sample and it is simple to estimate the new effective volume.  
However, it is important to use the correct minimum distance.  If it is too small, then 
the correction will not work for the largest groups; if it is too big, then we will 
unnecessarily reduce our sample size.

The left panel of Figure~\ref{fig:nbodyedgestats} shows the fraction of mock groups that 
are missing one or more galaxies due to edges, as a function of the distance from the
group centroid to the edge.  The fraction drops from 20\% at 100 Kpc to 5\% at 500 Kpc 
and less than 1\% at 1 Mpc.  It does not go to zero at larger distance because there are 
groups with high velocity dispersion that can be far from the edge and still have 
galaxies within a linking length of the outer or lower redshift limit of our sample.  
This figure suggests that if we set the minimum distance to 500 Kpc in the tangential 
direction and 500 km/s in the redshift direction, we should eliminate most groups that 
are affected by edges.  We make this correction on our mock group catalogs, and the 
number of groups in the resulting catalog is reduced by $\sim 22\%$ on average.  We 
estimate the new effective volume of each group catalog by scaling the original volume 
by the fraction of groups that survive the edge cut.  This estimate, though not exactly 
accurate, is simple to make and adequate for our purposes.  Figure \ref{fig:nbodyedge} 
shows that this correction results in a multiplicity function that is unbiased due to 
edges (dashed red curve).  

Our mock catalog tests show that we can deal with survey edges effectively if we measure
the multiplicity function after eliminating all groups whose centers (estimated as the
centroids of their member galaxy positions) lie less than 500 Kpc from an edge in the 
tangential direction or less than 500 km/s from an edge in the radial direction.  
Applying this edge cut to the $Mr20$, $Mr19$, and $Mr18$ SDSS group catalogs reduces the 
numbers of groups by 22.0\%, 30.2\%, and 41.1\%, respectively.  Our measurement of the 
multiplicity function for these samples includes this correction, though the group 
catalogs that we present include all groups.

%-----------------------------------------------------------------------
%-----------------------------------------------------------------------

%#############################################################################

\section{Group and Cluster Catalog} \label{catalog}

We apply our group-finding algorithm to the three volume-limited samples described
in \S~\ref{data} and get three group catalogs.  The fractions of ungrouped, isolated 
galaxies are 43.7\%, 41.2\%, and 39.8\% for the $Mr20$, $Mr19$, and $Mr18$ samples, 
respectively.  The fractions of galaxies grouped in pairs are 19.1\%, 18.3\%, and 
17.9\%.  The remaining 37.2\%, 40.6\%, and 42.3\% of galaxies are in groups of three or 
more members.  Samples $Mr20$, $Mr19$, and $Mr18$ contain a total of 4107, 2684, and 
1357 groups with richness $N\geq3$, respectively.

\begin{figure*}[p]
\epsscale{1.0}
\plotone{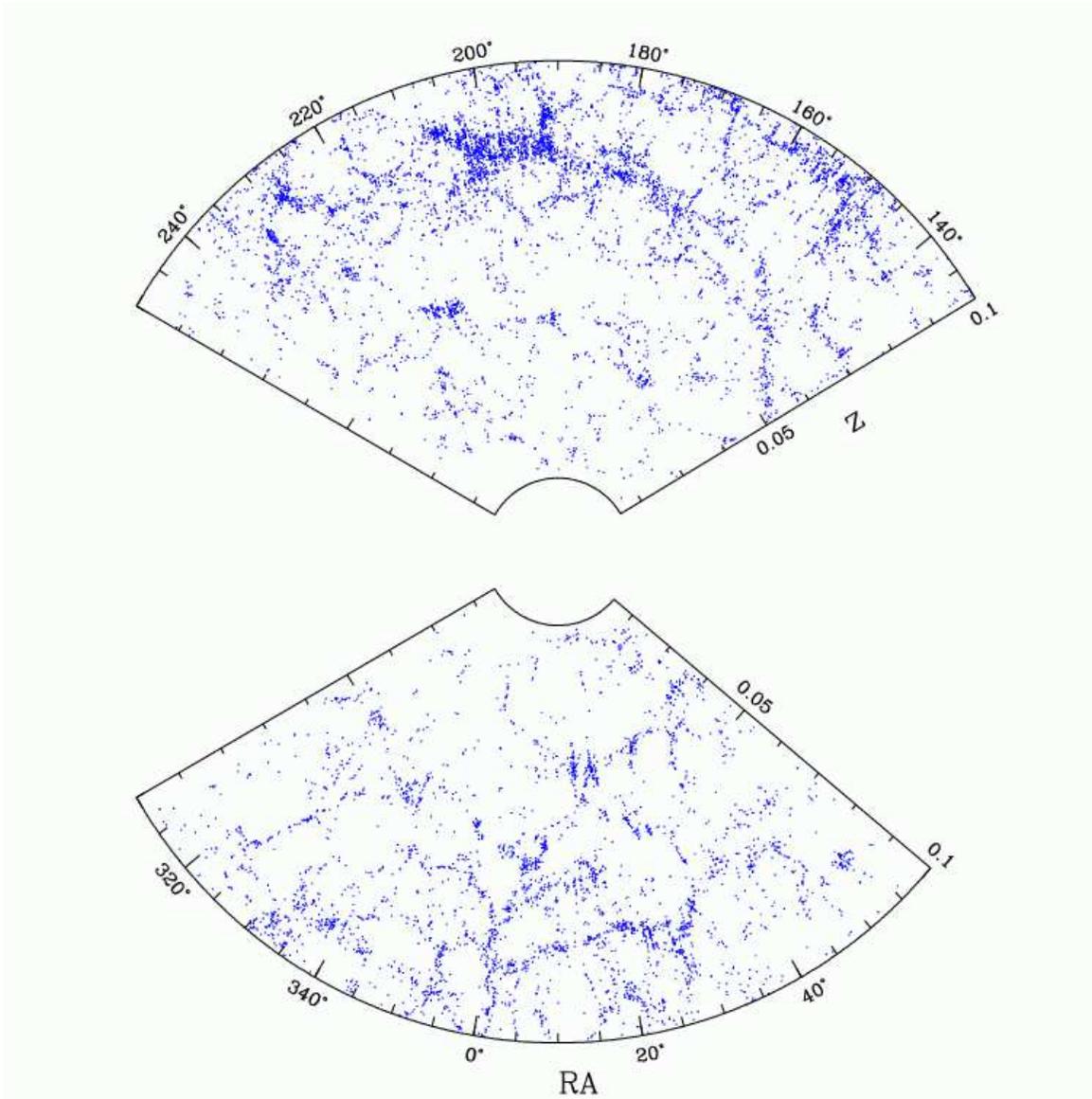}
\caption{
Equatorial slice through the SDSS volume-limited sample in the redshift 
range $0.015-0.1$.  The slice is $4^\circ$ thick and each point shows
the RA and redshift of a galaxy in the sample.
}
\label{fig:slicevollim}
\end{figure*}

\begin{figure*}[p]
\epsscale{1.0}
\plotone{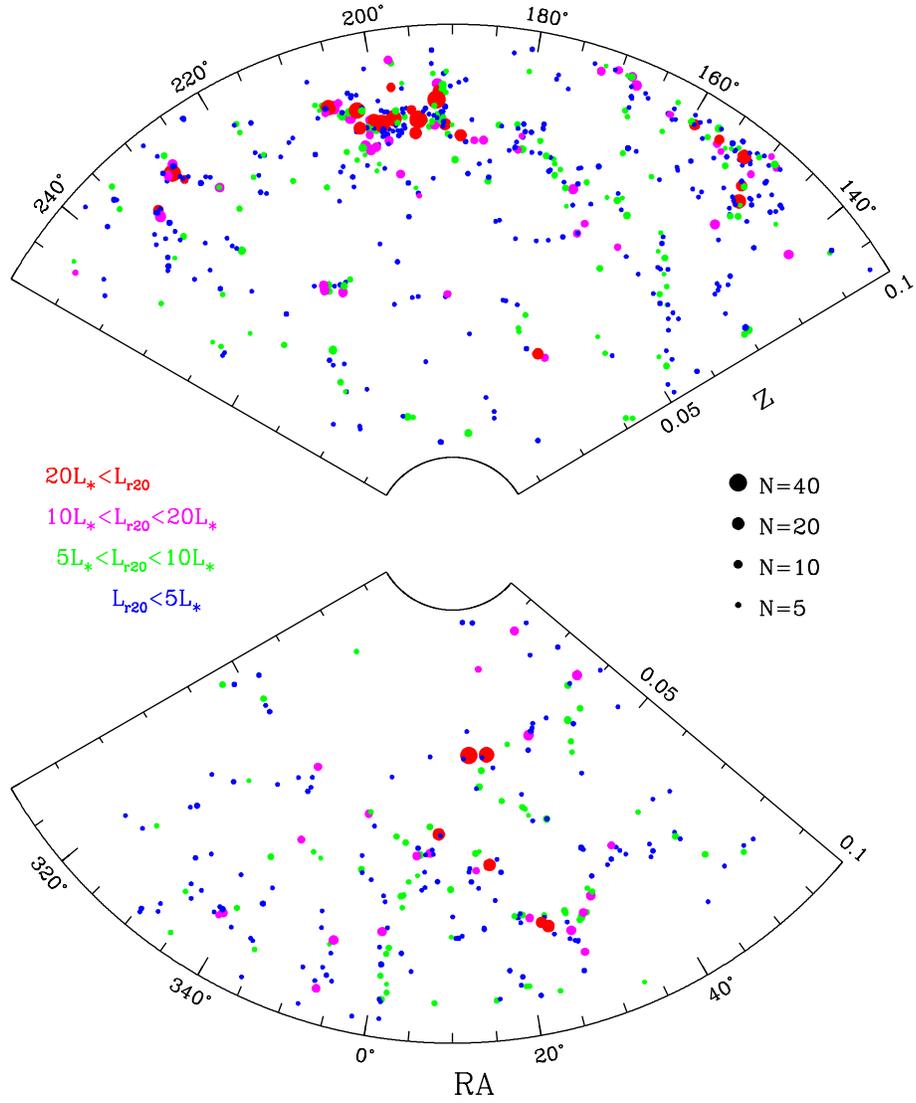}
\caption{
$4^\circ$ thick equatorial slice showing galaxy groups in the $Mr20$ 
volume-limited sample.  Each point shows the location of a group of richness 
$N\geq 3$.  Points have a size proportional to group richness $N$ and a color 
encoding according to their total $r$-band luminosity $\Lrtot$ (defined in the 
text) in units of $\Lstar$ (where we adopt $\Mstar=-20.44$), as listed in the 
legend.  
}
\label{fig:groupslice}
\end{figure*}

\begin{figure*}[p]
\epsscale{1.0}
\plotone{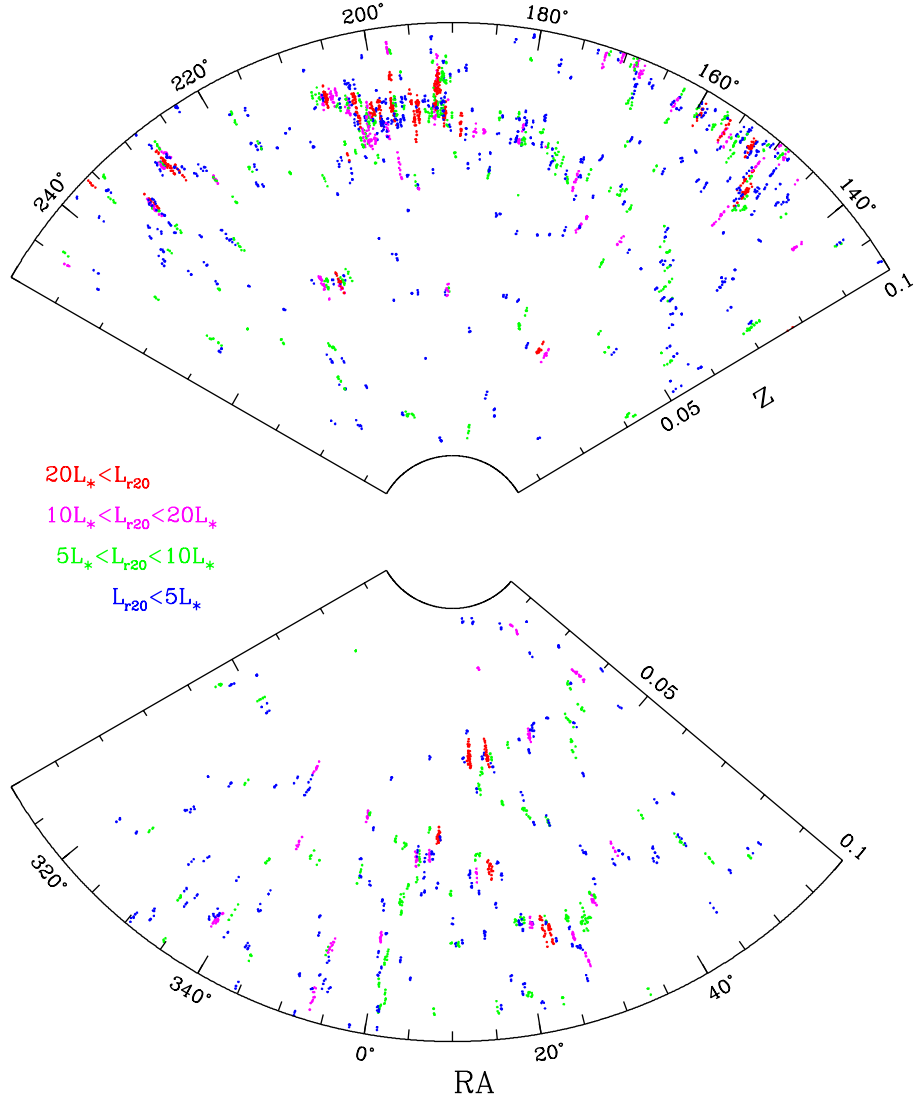}
\caption{
Same as Fig.~\ref{fig:groupslice} except that points show the locations of
member galaxies in groups of richness $N\geq 3$.
}
\label{fig:groupslice2}
\end{figure*}

Figure~\ref{fig:groupslice} shows an equatorial slice with groups identified from
sample $Mr20$.  The slice is $4^\circ$ thick and each point shows the RA and 
redshift of a group with $N\geq3$.  A comparison of this figure to 
Figure~\ref{fig:slicevollim} shows that groups and clusters trace the large-scale
structure of galaxies, as expected.  Larger groups are preferentially located
in higher density regions, whereas smaller groups are more uniformly distributed.
It is striking that the majority of very large groups reside within the large
supercluster at $z=0.08$.  Figure~\ref{fig:groupslice2} shows the same slice, but
with points representing the positions of member galaxies in $N\geq3$ groups.
A visual inspection of the figure shows that group velocity dispersions, which
are responsible for the finger-of-God effect, are largest in the most luminous
groups.

For each group, we compute an unweighted group centroid, which consists of a group 
right ascension, declination, and mean redshift.  We compute a total group 
luminosity that is the sum of luminosities of its member galaxies.  Since we are
dealing with volume-limited samples, the luminosity of a given group in samples
$Mr20$, $Mr19$, $Mr18$, only counts galaxies with absolute magnitudes brighter
than -19.9, -19, -18, respectively.  For example, for the $Mr20$ sample, the total
group absolute magnitude is
\begin{equation}
\Mrtot = -2.5 \mathrm{log}\left( \sum_{i=1}^{N} 10^{-0.4M_{\band{0.1}{r},i}} \right),
\end{equation}
and it is equivalent to integrating the galaxy luminosity function within the
group from $\Mr=-19.9$ to $-\infty$.  Note that we compute these group absolute 
magnitudes using the altered absolute magnitudes for galaxies that do not have
measured redshifts due to fiber collisions (see \S~\ref{data}).  
We also compute a total group color, which is simply defined as
$\grgrp = \Mgtot - \Mrtot$.  We compute a group one-dimensional velocity 
dispersion given by
\begin{equation}
\sigv = \frac{1}{1+\bar{z}}\sqrt{\frac{1}{N-1}\sum_{i=1}^{N} (cz_i - c\bar{z})^2},
\end{equation}
and an rms projected group radius given by
\begin{equation}
\Rproj = \sqrt{\frac{1}{N}\sum_{i=1}^{N} r_i^2},
\end{equation}
where $r_i$ is the projected distance between each member galaxy and the group
centroid.

In the three portions of Table~3, we present the groups and clusters with $N\geq3$,
selected from samples $Mr20$, $Mr19$, and $Mr18$.  For each group, we list a group
ID (column 1); the (J2000) right ascension and declination of the group centroid
(columns 2, 3); the mean redshift of the cluster (column 4); the group richness $N$
(column 5); the total $r$-band absolute magnitude of the group, $\Mrtot$ (column 6);
the total color of the group, $\grgrp$ (column 7); the line-of-sight velocity
dispersion of the group, $\sigv$ (column 8); the projected rms radius of the
group $\Rproj$ (column 9); the perpendicular distance of the group center from
the survey edge $\redge$ (column 10).  The groups in each portion of Table~3 are
ranked in decreasing order of richness $N$.  We show the first few rows of each
portion of the table in the text and make the entire table available in the electronic
version of the journal, as well as at \texttt{http://cosmo.nyu.edu/aberlind/Groups}.

In Table~4, we present the member galaxies of the groups listed in Table~3.  For
each galaxy we list the ID of the group to which it belongs (column 1); the (J2000)
right ascension and declination (columns 2, 3); the redshift (column 4); the
$r$-band absolute magnitude $\Mr$\footnote{Galaxies without measured redshifts due
to fiber collisions are assigned the absolute magnitude of their nearest neighbor,
as described in \S~\ref{data}.} (column 5); the $\gr$ color (column 6); a fiber
collision flag that is equal to 0 if the galaxy has its own measured redshift and 1
if it has been given the redshift of its nearest neighbor (column 7); the perpendicular
distance of the galaxy from the survey edge $\redge$ (column 8).  As before, we show
the first few rows of each portion of Table~4 in the text and make the entire table
available in the electronic version of the journal, as well as at
\texttt{http://cosmo.nyu.edu/aberlind/Groups}.

%-----------------------------------------------------------------------
%-----------------------------------------------------------------------
%-----------------------------------------------------------------------
%-----------------------------------------------------------------------
%-----------------------------------------------------------------------
%-----------------------------------------------------------------------
%-----------------------------------------------------------------------
\begin{table*}
\begin{center}
\centerline{\small Table~3. Group and Cluster Catalogs for Samples $Mr20$, $Mr19$, and $Mr18$}
\smallskip
\begin{tabular}{rrrccccccr}
\tableline
\tableline
ID & RA & DEC & $\bar{z}$ & $N$ & $\Mrtot$ & $\grgrp$ & $\sigv$ & $\Rproj$ & $\redge$ \\
 & (deg) & (deg) & & & & & (km/s) & ($\hmpc$) & ($\hmpc$) \\
\tableline
\cutinhead{$Mr20$}
33974 & 239.580740 &  27.312343 & 0.08797 & 132 & -25.920 & 0.946 &  723.7 & 1.371 &  17.7 \\
16089 & 247.172589 &  40.164633 & 0.03057 &  97 & -25.468 & 0.891 &  661.1 & 1.318 &  89.3 \\
 8817 & 358.535971 & -10.372017 & 0.07405 &  61 & -25.190 & 0.921 &  736.0 & 0.734 &  17.9 \\
14552 & 183.450292 &  59.266666 & 0.09386 &  51 & -24.861 & 0.808 &  338.3 & 1.079 &  22.9 \\
12289 & 159.824898 &   4.987457 & 0.06815 &  51 & -24.859 & 0.899 &  661.4 & 1.161 &  47.1 \\
 3025 & 195.700154 &  -2.627141 & 0.08183 &  49 & -24.805 & 0.911 &  377.1 & 1.247 &  57.9 \\
20593 & 169.362355 &  54.469262 & 0.06907 &  49 & -24.831 & 0.906 &  426.4 & 1.202 &  35.5 \\
\cutinhead{$Mr19$}
 9501 & 246.963120 &  40.182569 & 0.03009 & 197 & -25.839 & 0.886 &  588.7 & 1.317 &  88.2 \\
 4915 &  10.447791 &  -9.381301 & 0.05543 &  95 & -25.068 & 0.927 &  572.4 & 0.981 &  38.8 \\
 4634 & 329.333792 &  -7.765802 & 0.05727 &  86 & -25.016 & 0.724 &  564.0 & 0.677 &  52.5 \\
10986 &  14.231949 &  -0.655097 & 0.04378 &  86 & -24.944 & 0.935 &  385.4 & 1.076 &   5.2 \\
 5585 & 351.303515 &  14.909898 & 0.04113 &  83 & -24.622 & 0.871 &  496.8 & 1.045 &  53.2 \\
 3709 & 214.187113 &   1.962572 & 0.05333 &  81 & -24.902 & 0.887 &  368.3 & 1.160 &  42.9 \\
11585 &  18.686704 &   0.254973 & 0.04442 &  68 & -24.704 & 0.903 &  386.8 & 0.744 &  27.0 \\
\cutinhead{$Mr18$}
 4792 & 247.062059 &  40.107520 & 0.03011 & 311 & -25.934 & 0.865 &  584.2 & 1.300 &  90.5 \\
 2748 & 351.183638 &  14.580962 & 0.04128 & 152 & -25.057 & 0.903 &  446.6 & 1.014 &  72.3 \\
 6984 & 173.640705 &  49.042739 & 0.03270 &  65 & -24.086 & 0.918 &  526.2 & 0.533 &  45.7 \\
 1968 & 220.146510 &   3.491413 & 0.02680 &  54 & -23.853 & 0.946 &  274.1 & 0.506 &  23.6 \\
 5607 &  14.274495 &  -0.247149 & 0.04303 &  52 & -24.066 & 0.915 &  309.0 & 0.760 &  13.0 \\
 5948 &  18.760997 &   0.307893 & 0.04326 &  49 & -24.108 & 0.876 &  264.9 & 0.659 &  26.5 \\
 5692 &  51.279369 &  -0.496506 & 0.03664 &  48 & -23.871 & 0.870 &  246.1 & 0.802 &  44.6 \\
\tableline
\label{tab:groups}
\end{tabular}
\end{center}
Note---The rest of the table can be found in the electronic version of the ApJ, or at 
\texttt{http://cosmo.nyu.edu/aberlind/Groups}
\end{table*}
%-----------------------------------------------------------------------
%-----------------------------------------------------------------------
\begin{table*}
\begin{center}
\centerline{\small Table~4. Member Galaxies of Groups and Clusters for Samples $Mr20$, $Mr19$, and $Mr18$}
\smallskip
\begin{tabular}{lcrccccc}
\tableline
\tableline
groupID & RA & DEC & $z$ & $\Mr$ & $\gr$ & fibcol & $\redge$ \\
 & (deg) & (deg) & & & & & ($\hmpc$) \\
\tableline
\cutinhead{$Mr20$}
   14 & 196.769894 &  -0.039161 & 0.08086 & -20.168 & 0.945 & 1 &  72.3 \\
   14 & 196.799107 &  -0.024688 & 0.08051 & -20.498 & 0.918 & 0 &  72.3 \\
   14 & 196.788454 &  -0.029741 & 0.08086 & -20.168 & 0.945 & 1 &  72.3 \\
   14 & 196.779246 &  -0.038656 & 0.08086 & -20.168 & 0.945 & 0 &  72.3 \\
   15 & 197.264020 &  -0.053520 & 0.07962 & -20.302 & 0.457 & 0 &  72.4 \\
   15 & 197.207327 &   0.047123 & 0.07987 & -19.950 & 0.895 & 0 &  72.4 \\
   15 & 197.165432 &   0.102322 & 0.08016 & -20.467 & 0.872 & 0 &  72.4 \\
\cutinhead{$Mr19$}
    1 & 169.180550 &  -0.213320 & 0.03917 & -19.355 & 0.752 & 0 &  13.5 \\
    1 & 169.195964 &  -0.100215 & 0.03898 & -19.315 & 0.584 & 0 &  13.5 \\
    1 & 169.387065 &  -0.187503 & 0.03999 & -20.762 & 0.967 & 0 &  13.5 \\
    5 & 199.555960 &  -0.148218 & 0.04825 & -19.267 & 0.321 & 0 &  65.9 \\
    5 & 199.656619 &  -0.226944 & 0.04731 & -19.705 & 0.960 & 0 &  65.9 \\
    5 & 199.665084 &  -0.175183 & 0.04708 & -20.975 & 0.976 & 1 &  65.9 \\
    5 & 199.679052 &  -0.178932 & 0.04708 & -20.975 & 0.976 & 0 &  65.9 \\
    5 & 199.671638 &  -0.173772 & 0.04708 & -20.975 & 0.976 & 1 &  65.9 \\
\cutinhead{$Mr18$}
    1 & 194.342587 &  -0.630508 & 0.02247 & -18.821 & 0.744 & 1 &  57.7 \\
    1 & 194.353591 &  -0.622488 & 0.02247 & -18.821 & 0.744 & 0 &  57.7 \\
    1 & 194.313130 &  -0.657646 & 0.02295 & -18.837 & 0.894 & 0 &  57.7 \\
    2 & 169.180550 &  -0.213320 & 0.03917 & -19.355 & 0.752 & 0 &  13.4 \\
    2 & 169.195964 &  -0.100215 & 0.03898 & -19.315 & 0.584 & 0 &  13.4 \\
    2 & 169.387065 &  -0.187503 & 0.03999 & -20.762 & 0.967 & 0 &  13.4 \\
    2 & 169.300864 &  -0.189302 & 0.03972 & -18.203 & 0.819 & 0 &  13.4 \\
\tableline
\label{tab:members}
\end{tabular}
\end{center}
Note---The rest of the table can be found in the electronic version of the ApJ, or at 
\texttt{http://cosmo.nyu.edu/aberlind/Groups}
\end{table*}
%-----------------------------------------------------------------------
\clearpage

%#############################################################################

\section{Multiplicity Function} \label{multiplicity}

\begin{figure}[t]
\epsscale{1.0}
\plotone{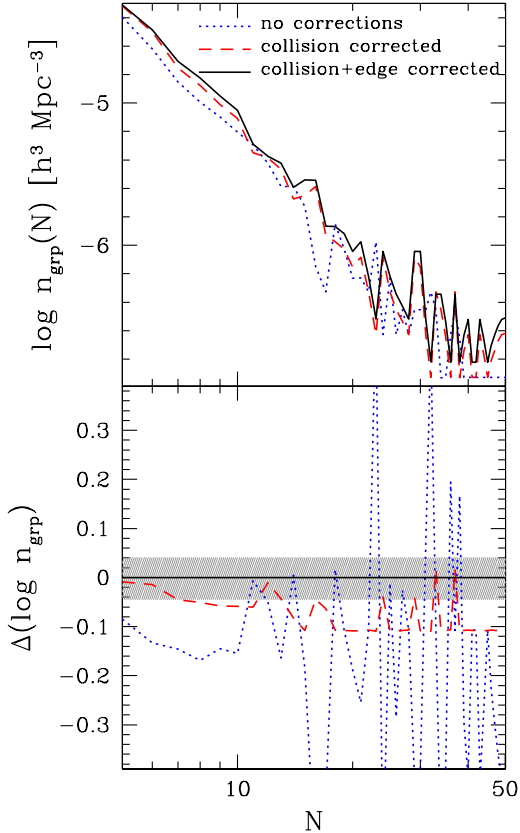}
\caption{
Differential group multiplicity function for groups identified in the SDSS
$Mr20$ volume-limited sample.  The different curves are $\ngrpN$ uncorrected
for incompleteness ({\it dotted blue curves}), corrected for incompleteness
due to fiber collisions ({\it dashed red curves}), and corrected for both fiber
collisions and edge effects ({\it solid black curves}).  The bottom panel shows the
ratio of each case to the fully corrected one.  The shaded region encloses
$\pm 10\%$ deviations from the fully corrected multiplicity function.
These results are averaged over all of our \texttt{.Mr20} mock catalogs.
}
\label{fig:groupmultfibedge}
\end{figure}

With group catalogs in hand, we can now measure the group multiplicity function.
The differential group multiplicity function, $\ngrpN$, is defined as the
number density of groups in bins of richness $N$, where richness bins can have
a width of unity or more.  Before computing $\ngrpN$, we must make the
corrections for incompleteness described in \S~\ref{incompleteness}.  
Though the catalogs presented in \S~\ref{catalog} already include the fiber
collision correction, we also compute the multiplicity function from an
alternate $Mr20$ group catalog that does not include this correction in order
to see the magnitude of the correction.  Figure~\ref{fig:groupmultfibedge}
shows this uncorrected multiplicity function, as well as the multiplicity
function that includes the fiber collision correction.  The figure shows that
applying the correction boosts the amplitude of the multiplicity function, just
as it did in our mock tests in \S~\ref{incompleteness}.
Figure~\ref{fig:groupmultfibedge} also shows the effect on the multiplicity
function of applying the edge correction described in \S~\ref{incompleteness}.
This effect is small, typically less than 5\%, though it is larger in individual bins 
at high $N$, where the number of groups is small.

%-----------------------------------------------------------------------
%-----------------------------------------------------------------------
%-----------------------------------------------------------------------
\begin{table}[t]
\begin{center}
\centerline{\small Table~5. Group Multiplicity Function for $Mr20$ Sample}
\begin{tabular}[t]{lccc}
\tableline
\tableline
$\Nmin$--$\Nmax$ & $\ngrpN$ & $\signgrpN$ & $\signgrpN$ (Poisson) \\
\tableline
 3--3 & $2.290\times 10^{-4}$ & $1.110\times 10^{-5}$ & $5.881\times 10^{-6}$ \\
 4--4 & $1.054\times 10^{-4}$ & $4.890\times 10^{-6}$ & $3.990\times 10^{-6}$ \\
 5--5 & $4.909\times 10^{-5}$ & $4.181\times 10^{-6}$ & $2.723\times 10^{-6}$ \\
 6--6 & $3.263\times 10^{-5}$ & $4.465\times 10^{-6}$ & $2.220\times 10^{-6}$ \\
 7--7 & $1.962\times 10^{-5}$ & $1.979\times 10^{-6}$ & $1.722\times 10^{-6}$ \\
 8--8 & $1.496\times 10^{-5}$ & $2.250\times 10^{-6}$ & $1.503\times 10^{-6}$ \\
 9--9 & $1.118\times 10^{-5}$ & $2.398\times 10^{-6}$ & $1.299\times 10^{-6}$ \\
10--10 & $8.906\times 10^{-6}$ & $1.502\times 10^{-6}$ & $1.160\times 10^{-6}$ \\
11--11 & $5.139\times 10^{-6}$ & $1.292\times 10^{-6}$ & $8.810\times 10^{-7}$ \\
12--12 & $4.223\times 10^{-6}$ & $8.632\times 10^{-7}$ & $7.986\times 10^{-7}$ \\
13--13 & $3.780\times 10^{-6}$ & $7.200\times 10^{-7}$ & $7.555\times 10^{-7}$ \\
14--14 & $2.565\times 10^{-6}$ & $1.283\times 10^{-6}$ & $6.224\times 10^{-7}$ \\
15--15 & $2.873\times 10^{-6}$ & $9.335\times 10^{-7}$ & $6.587\times 10^{-7}$ \\
16--16 & $2.868\times 10^{-6}$ & $1.165\times 10^{-6}$ & $6.581\times 10^{-7}$ \\
17--17 & $1.361\times 10^{-6}$ & $6.868\times 10^{-7}$ & $4.533\times 10^{-7}$ \\
18--18 & $1.358\times 10^{-6}$ & $4.131\times 10^{-7}$ & $4.530\times 10^{-7}$ \\
19--19 & $1.209\times 10^{-6}$ & $5.133\times 10^{-7}$ & $4.273\times 10^{-7}$ \\
20--21 & $9.817\times 10^{-7}$ & $3.079\times 10^{-7}$ & $3.851\times 10^{-7}$ \\
22--24 & $6.039\times 10^{-7}$ & $2.253\times 10^{-7}$ & $3.020\times 10^{-7}$ \\
25--28 & $3.401\times 10^{-7}$ & $9.522\times 10^{-8}$ & $2.266\times 10^{-7}$ \\
29--30 & $9.061\times 10^{-7}$ & $4.483\times 10^{-7}$ & $3.699\times 10^{-7}$ \\
31--34 & $3.398\times 10^{-7}$ & $7.501\times 10^{-8}$ & $2.265\times 10^{-7}$ \\
35--42 & $1.699\times 10^{-7}$ & $6.455\times 10^{-8}$ & $1.602\times 10^{-7}$ \\
43--61 & $6.360\times 10^{-8}$ & $2.982\times 10^{-8}$ & $9.801\times 10^{-8}$ \\
\tableline
\label{tab:mult20}
\end{tabular}
\end{center}
Note---$\ngrp$ and $\signgrpN$ are in units of $\hden$.
\end{table}
%-----------------------------------------------------------------------

We must calculate errorbars for the multiplicity function in order to use it to
constrain the HOD.  We use our mock catalogs for this purpose.  Specifically,
we compute fractional errors from the dispersion among 10 independent mock catalogs for 
the $Mr20$ sample (\texttt{LANL1-5.Mr20} mocks $\times$ 1 HOD realization $\times$ 
2 mocks per simulation cube), and 8 mock catalogs for each of the $Mr19$ and $Mr18$
samples (\texttt{LANL1-4.Mr19}/\texttt{LANL1-4.Mr18} mocks $\times$ 1 HOD 
realization $\times$ 2 mocks per simulation cube).  Note that we do not use multiple 
HOD realizations because the underlying halo populations themselves would not be 
independent.  Before computing errors, we correct each mock catalog for fiber 
collisions and edge effects in the same way as in the data.  The computed errors
thus implicitly include any contribution from these correction procedures.

The SDSS multiplicity function shown in Figure~\ref{fig:groupmultfibedge} becomes
very noisy at high richness because the abundance of groups drops with $N$ and
the figure uses richness bins with a width of unity.  It makes more sense to
increase the bin width with $N$ so as to beat down the noise.  Moreover, since
we calculate errorbars for the multiplicity function using our mock catalogs,
each richness bin must contain enough mock groups so that an errorbar can be
reliably estimated.  We choose richness bins for each group catalog so that each
bin contains at least eight SDSS groups and twenty mock groups (among all mock
catalogs used).  At low multiplicities, the bin width is always unity because
there are many groups with low $N$.  At higher multiplicities, however, the
richness bins grow wider in order to satisfy these criteria.  The bin widths
for samples $Mr20$, $Mr19$, and $Mr18$, are listed in the first columns of Tables~5,
6, and~7, respectively.  Once a richness bin is defined, the abundance of groups in
that bin, $\ngrpN$, is simply the number of groups having richnesses within the bin,
divided by the sample volume and divided by the bin width.  The values of $\ngrpN$ are
listed in the second columns of Tables~5, 6, and~7.  We use the same richness bins
to compute the abundance of mock groups for each independent mock catalog, and we
compute errors, $\signgrpN$, in the SDSS multiplicity function by measuring the
dispersion among the mock multiplicity functions.  These errors are listed in the
third columns of Tables~5, 6, and~7.  Finally, we also compute Poisson errors for
the SDSS $\ngrpN$, which we list in the fourth columns of Tables~5, 6, and~7.
In some of the highest multiplicity bins, the Poisson errors are larger than the mock
errors.  In these cases, the mock errors are likely underestimated and it is
best to use the Poisson errors in their place.
%-----------------------------------------------------------------------
\begin{table}[t]
\begin{center}
\centerline{\small Table~6. Group Multiplicity Function for $Mr19$ Sample}
\begin{tabular}[t]{lccc}
\tableline
\tableline
$\Nmin$--$\Nmax$ & $\ngrpN$ & $\signgrpN$ & $\signgrpN$ (Poisson) \\
\tableline
 3--3 & $4.514\times 10^{-4}$ & $2.872\times 10^{-5}$ & $1.545\times 10^{-5}$ \\
 4--4 & $1.889\times 10^{-4}$ & $1.201\times 10^{-5}$ & $9.996\times 10^{-6}$ \\
 5--5 & $1.085\times 10^{-4}$ & $9.323\times 10^{-6}$ & $7.575\times 10^{-6}$ \\
 6--6 & $6.292\times 10^{-5}$ & $8.977\times 10^{-6}$ & $5.769\times 10^{-6}$ \\
 7--7 & $5.027\times 10^{-5}$ & $5.465\times 10^{-6}$ & $5.157\times 10^{-6}$ \\
 8--8 & $2.856\times 10^{-5}$ & $2.434\times 10^{-6}$ & $3.887\times 10^{-6}$ \\
 9--9 & $1.853\times 10^{-5}$ & $2.832\times 10^{-6}$ & $3.131\times 10^{-6}$ \\
10--10 & $1.534\times 10^{-5}$ & $2.799\times 10^{-6}$ & $2.849\times 10^{-6}$ \\
11--11 & $1.534\times 10^{-5}$ & $2.577\times 10^{-6}$ & $2.849\times 10^{-6}$ \\
12--12 & $1.164\times 10^{-5}$ & $2.236\times 10^{-6}$ & $2.482\times 10^{-6}$ \\
13--13 & $8.994\times 10^{-6}$ & $2.135\times 10^{-6}$ & $2.181\times 10^{-6}$ \\
14--14 & $7.936\times 10^{-6}$ & $2.105\times 10^{-6}$ & $2.049\times 10^{-6}$ \\
15--15 & $5.819\times 10^{-6}$ & $1.186\times 10^{-6}$ & $1.755\times 10^{-6}$ \\
16--16 & $5.819\times 10^{-6}$ & $1.718\times 10^{-6}$ & $1.755\times 10^{-6}$ \\
17--18 & $5.819\times 10^{-6}$ & $1.318\times 10^{-6}$ & $1.755\times 10^{-6}$ \\
19--20 & $2.380\times 10^{-6}$ & $5.168\times 10^{-7}$ & $1.122\times 10^{-6}$ \\
21--23 & $2.292\times 10^{-6}$ & $5.243\times 10^{-7}$ & $1.101\times 10^{-6}$ \\
24--26 & $1.587\times 10^{-6}$ & $4.621\times 10^{-7}$ & $9.164\times 10^{-7}$ \\
27--32 & $7.054\times 10^{-7}$ & $2.228\times 10^{-7}$ & $6.109\times 10^{-7}$ \\
33--38 & $7.054\times 10^{-7}$ & $3.069\times 10^{-7}$ & $6.109\times 10^{-7}$ \\
39--51 & $3.256\times 10^{-7}$ & $4.634\times 10^{-8}$ & $4.151\times 10^{-7}$ \\
52--86 & $1.209\times 10^{-7}$ & $3.602\times 10^{-8}$ & $2.529\times 10^{-7}$ \\
\tableline
\label{tab:mult19}
\end{tabular}
\end{center}
Note---Same units as Table~5.
\end{table}
%-----------------------------------------------------------------------
%-----------------------------------------------------------------------
\begin{table}[t]
\begin{center}
\centerline{\small Table~7. Group Multiplicity Function for $Mr18$ Sample}
\begin{tabular}[t]{lccc}
\tableline
\tableline
$\Nmin$--$\Nmax$ & $\ngrpN$ & $\signgrpN$ & $\signgrpN$ (Poisson) \\
\tableline
 3--3 & $7.311\times 10^{-4}$ & $6.909\times 10^{-5}$ & $4.000\times 10^{-5}$ \\
 4--4 & $3.436\times 10^{-4}$ & $3.325\times 10^{-5}$ & $2.742\times 10^{-5}$ \\
 5--5 & $1.948\times 10^{-4}$ & $2.200\times 10^{-5}$ & $2.065\times 10^{-5}$ \\
 6--6 & $1.248\times 10^{-4}$ & $1.629\times 10^{-5}$ & $1.652\times 10^{-5}$ \\
 7--7 & $1.182\times 10^{-4}$ & $1.546\times 10^{-5}$ & $1.608\times 10^{-5}$ \\
 8--8 & $5.686\times 10^{-5}$ & $9.917\times 10^{-6}$ & $1.116\times 10^{-5}$ \\
 9--9 & $3.284\times 10^{-5}$ & $5.340\times 10^{-6}$ & $8.477\times 10^{-6}$ \\
10--10 & $3.066\times 10^{-5}$ & $5.777\times 10^{-6}$ & $8.191\times 10^{-6}$ \\
11--11 & $2.626\times 10^{-5}$ & $8.403\times 10^{-6}$ & $7.581\times 10^{-6}$ \\
12--13 & $1.423\times 10^{-5}$ & $1.629\times 10^{-6}$ & $5.580\times 10^{-6}$ \\
14--15 & $8.756\times 10^{-6}$ & $1.443\times 10^{-6}$ & $4.378\times 10^{-6}$ \\
16--17 & $1.203\times 10^{-5}$ & $1.761\times 10^{-6}$ & $5.132\times 10^{-6}$ \\
18--23 & $3.647\times 10^{-6}$ & $7.402\times 10^{-7}$ & $2.825\times 10^{-6}$ \\
24--31 & $2.188\times 10^{-6}$ & $6.091\times 10^{-7}$ & $2.188\times 10^{-6}$ \\
32--152 & $1.447\times 10^{-7}$ & $1.673\times 10^{-8}$ & $5.627\times 10^{-7}$ \\
\tableline
\label{tab:mult18}
\end{tabular}
\end{center}
Note---Same units as Table~5.
\end{table}
%-----------------------------------------------------------------------

%-----------------------------------------------------------------------
\begin{figure}[t]
\epsscale{1.0}
\plotone{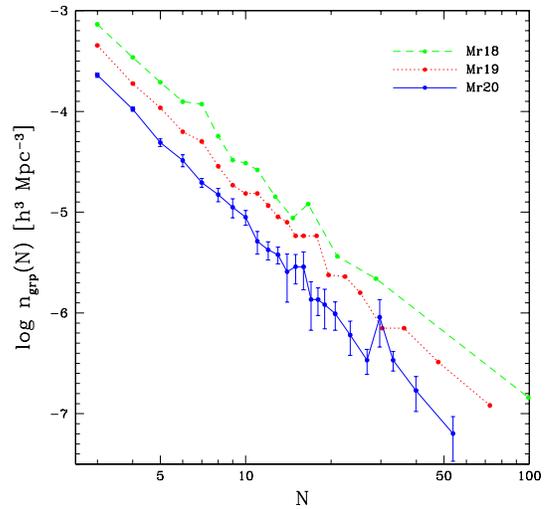}
\caption{
Differential group multiplicity functions for SDSS groups.  The three curves
show $\ngrpN$ for groups identified in our three volume-limited samples:
$Mr20$, $Mr19$, and $Mr18$ (colors and line types are listed in the top-right
corner of the panel).  $\ngrpN$ is measured in richness bins whose widths are
chosen so that the bins contain a minimum of 8 SDSS groups and 20 mock groups.
Points are placed at the mean richness of groups within each bin.
Errors are shown for the $Mr20$ sample and are estimated from the dispersion
among 10 independent SDSS mock catalogs.
}
\label{fig:groupmult}
\end{figure}
%-----------------------------------------------------------------------

Figure~\ref{fig:groupmult} shows the SDSS multiplicity functions for the three
volume-limited samples, along with the mock errorbars for the $Mr20$ sample.
Though we measure and show the multiplicity function down to a multiplicity of
$N=3$, our tests with mock catalogs have shown that it is only unbiased with
respect to the true halo multiplicity function for $N\geq 10$.
When using this measured multiplicity function to constrain the HOD, we must
either only use bins with $N\geq 10$, or attempt to calibrate the relation between
the measured group multiplicity function and the true halo multiplicity function
at lower values of $N$. The central curve of Figure~\ref{fig:nbodymultbxybz}, 
discussed in the Appendix, effectively provides this calibration for $Mr20$ and the 
cosmology adopted in our mock catalogs.

The multiplicity functions shown in Figure~\ref{fig:groupmult} appear to be close
to power-law relations.  In order to test this, we perform a simple power-law fit
to each multiplicity function in the regime $N\geq 10$.  We use only the diagonal
errors of the full covariance matrix (i.e., the errors listed in Tables~5, 6,
and ~7).  We find that all three multiplicity functions are well-fit by power-law
relations, with best-fit slopes of $-2.72\pm0.16$, $-2.48\pm0.14$, and $-2.49\pm0.28$
for the $Mr20$, $Mr19$, and $Mr18$ samples, respectively.

%#############################################################################

\section{Summary and Discussion} \label{summary}

We have used a simple friends-of-friends algorithm to identify galaxy groups in
volume-limited samples of the SDSS redshift survey.  We have selected FoF
linking lengths that are best at grouping together galaxies that occupy the same 
dark matter halos.  We based this choice on extensive tests with mock galaxy 
catalogs, which we constructed by populating halos in N-body simulations with 
galaxies.  The result of our mock tests is that no combination of perpendicular
and line-of-sight linking lengths can yield groups that successfully recover 
all aspects of the parent halo distribution, even for large richness systems.
Specifically, FoF cannot identify groups that simultaneously have unbiased 
abundances, projected sizes, and velocity dispersions.  The ideal group-finding 
parameters for a given study depend on its scientific objectives.  Given our 
objective of using the multiplicity function to constrain the HOD, it makes
sense to sacrifice velocity dispersions and obtain groups with unbiased abundances
and projected sizes.  Our choice of linking lengths results in a group catalog
that, for groups of ten or more members, has an unbiased multiplicity function,
an unbiased median relation between the multiplicities of groups and their parent
halos, an unbiased projected size distribution as a function of multiplicity, and
a velocity dispersion distribution that is $\sim 20\%$ too low for all 
multiplicities.  We correct for fiber collisions and survey edge effects and present 
three SDSS group catalogs (for three different volume-limited samples) and their 
measured multiplicity functions.

It is important to recognize that our adopted group finder has the above properties 
only for halos defined using FoF with a linking length of 0.2 times the mean 
interparticle separation, since this is how halos were identified in our mock
catalogs.  A different halo definition (such as FoF with a different linking 
length, or spherical overdensity halos) would require a different set of optimal 
group-finding parameters.  This is not a problem as long as the same halo 
definition is used consistently.  For example, an HOD measured from these
group catalogs will hold for this halo definition, and any theoretical model should 
use the same halo definition to compare its predictions to the measured HOD.
We chose this particular halo finder because it has been widely used and tested,
and the properties of the resulting halo distribution (e.g., mass function) are
well understood.

The groups and clusters that we present here are intended to be systems of galaxies
that belong to the same virialized dark matter halo.  We can test whether these
systems are virialized by computing crossing times for the groups and checking
if they are sufficiently less than the Hubble time.  We define the crossing time
divided by the hubble time as 
\begin{equation}
\frac{\tcross}{\tH} = \frac{(\Rrms/\hmpc)}{(\sigv/100\kms)},
\end{equation}
where $\Rrms$ is the one-dimensional group radius, which is equal to the projected
(two-dimensional) radius, $\Rproj$, divided by the square root of two.  We correct
for the velocity dispersion bias revealed in our mock tests by applying a 20\% upward 
correction to all group velocity dispersions, and we compute $\tcross/\tH$ for all 
groups.  We find that, for all three group catalogs, the median value of 
$\tcross/\tH$ is $\sim0.15$, and 80\% of all groups have values less than $\sim0.29$.
These numbers can be interpreted in terms of the spherical infall model
\citep{gunn_gott_72,gott_turner_77a}, or other analytic or numerical models.
However, at a first glance, the numbers are encouraging and suggest that most of our 
groups are likely virialized systems.

The group and cluster catalogs presented here are well-suited for testing many of the 
predictions and assumptions made by galaxy formation models regarding the relationship 
between galaxies and their underlying dark matter halos.  We will investigate several 
of these issues in subsequent papers.

%#############################################################################

\acknowledgments 

We thank Zheng Zheng and Jeremy Tinker for their help with choosing HOD parameters for 
constructing mock catalogs and Luis Teodoro for his help with making the mock catalogs.  

AAB acknowledges support by NSF grant AST-0079251 and the NSF Center for Cosmological 
Physics, while at the University of Chicago, and by NASA grant NAG5-11669, NSF grant 
PHY-0101738, and a grant from NASA administered by the American Astronomical Society,
while at New York University.  AAB also acknowledges the hospitality of the Aspen
Center for Physics, where some of this work was completed.  MRB and DWH acknowledge
support by NSF grant AST-0428465.  DHW acknowledges support by NSF grant AST-0407125.  
JRG acknowledges support by NSF grant AST-0406713.
Portions of this work were performed under the auspices of the U.S. Dept. of Energy, 
and supported by its contract \#W-7405-ENG-36 to Los Alamos National Laboratory.

Funding for the SDSS and SDSS-II has been provided by the Alfred P. Sloan Foundation, the Participating Institutions, the National Science Foundation, the U.S. Department of Energy, the National Aeronautics and Space Administration, the Japanese Monbukagakusho, the Max Planck Society, and the Higher Education Funding Council for England. The SDSS Web Site is http://www.sdss.org/.

The SDSS is managed by the Astrophysical Research Consortium for the Participating Institutions. The Participating Institutions are the American Museum of Natural History, Astrophysical Institute Potsdam, University of Basel, Cambridge University, Case Western Reserve University, University of Chicago, Drexel University, Fermilab, the Institute for Advanced Study, the Japan Participation Group, Johns Hopkins University, the Joint Institute for Nuclear Astrophysics, the Kavli Institute for Particle Astrophysics and Cosmology, the Korean Scientist Group, the Chinese Academy of Sciences (LAMOST), Los Alamos National Laboratory, the Max-Planck-Institute for Astronomy (MPA), the Max-Planck-Institute for Astrophysics (MPIA), New Mexico State University, Ohio State University, University of Pittsburgh, University of Portsmouth, Princeton University, the United States Naval Observatory, and the University of Washington.

%#############################################################################

%\appendix \label{appendix}
\section*{Appendix}

In this appendix, we describe the mock catalog tests that help us choose optimal
FoF parameters.  Since our primary goal for identifying groups is to measure the 
group multiplicity function and use it to constrain the HOD, we clearly require
our FoF algorithm to produce groups that have an unbiased multiplicity function 
with respect to the true halo multiplicity function.  In addition, we require an
unbiased relation between the multiplicities of groups and their associated halos.
Finally, we would like our groups to have unbiased projected size and velocity 
dispersion distributions as a function of multiplicity.  We create a grid of
FoF linking lengths and check how each set of linking lengths performs in the 
above tests, for each of the four HOD model mock cubes 
(\texttt{.Mr20, .Mr20b, .Mr19, .Mr18}).  In the case of each HOD model, we average 
results over the 10 HOD realizations described in \S~\ref{mocks} and over the 
\texttt{LANL1} and \texttt{LANL4} N-body simulations.

\begin{figure}[t]
\epsscale{1.0}
\plotone{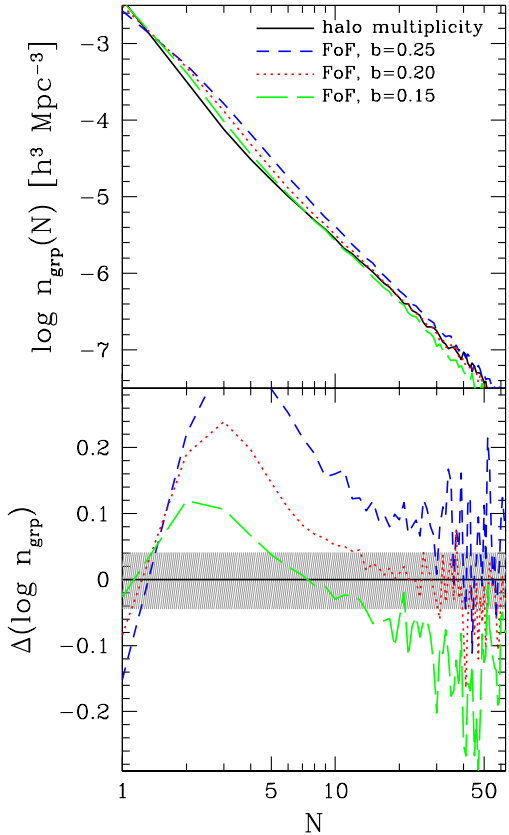}
\caption{
Effect of changing the FoF linking length on the group multiplicity function
in real space, measured using mock galaxy catalogs (described in \S~\ref{mocks}).
In the top panel, the solid black curve shows the input halo multiplicity function
for mock catalogs and thus represents the ``true'' case.  The other three curves
show the recovered group multiplicity functions for three different linking
lengths, which are listed at the top right of the panel in units of the mean
inter-galaxy separation.  The bottom panel shows the ratio of each case to the
``true'' one.  The shaded region encloses $\pm 10\%$ deviations from the ``true''
multiplicity function.
These results are averaged over all of our \texttt{.Mr20} mock catalogs.
}
\label{fig:nbodymultbxy}
\end{figure}

Before focusing on redshift space, we briefly examine how well FoF recovers
the true multiplicity function in real space, since this represents the best
possible case (any group finder will almost certainly perform worse in redshift 
space).  We apply FoF to the real-space cube mocks using a single linking length
(the linking volume around each mock galaxy is a sphere), and investigate how
the recovered multiplicity function varies with the value of this linking length.
In particular, we compare the mock group multiplicity functions to the input
halo multiplicity functions that were used to construct the mock catalogs.  
Figure~\ref{fig:nbodymultbxy} shows this comparison for the \texttt{.Mr20} mocks.
The bottom panel of the figure shows the logarithm of the ratio of group to halo 
multiplicity function, and the horizontal solid line therefore denotes the
``unbiased'' case.  The figure reveals that, at large $N$, the group multiplicity 
function has an unbiased shape that is independent of the choice of linking length
(at least for the range of linking lengths shown).  The amplitude, however, is 
dependent on the linking length used, with larger linking lengths leading to a 
higher abundance of groups at large $N$.  A linking length of $b=0.2$ (in units of
the mean intergalaxy separation) yields a group multiplicity function with an unbiased 
amplitude at large $N$.  This is not surprising given that the same value was used
to identify dark matter halos in the N-body simulations while constructing mock
catalogs.

%-----------------------------------------------------------------------
%-----------------------------------------------------------------------
%-----------------------------------------------------------------------
\begin{figure*}[p]
\epsscale{1.0}
\plotone{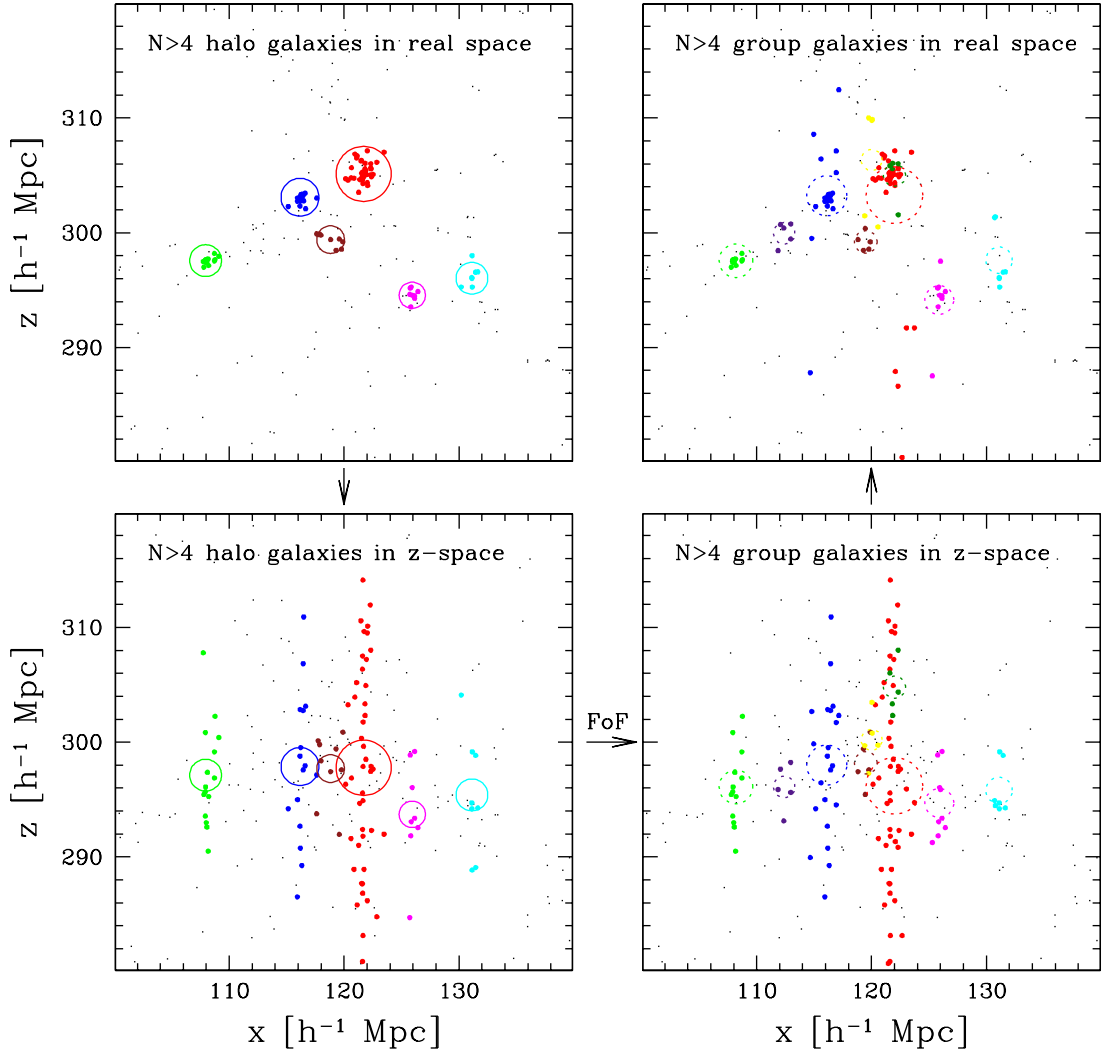}
\caption{
Illustrated behavior of the Friends-of-Friends (FoF) group finder.  Each panel 
shows a $40\times40\times10 \hmpc$ slice through a mock galaxy catalog.
Moving counter-clockwise starting from the top left panel, the panels show:
galaxies in dark matter halos in real space ({\it top left}), the same galaxies 
in redshift space ({\it bottom left}), galaxies in groups recovered using FoF 
({\it bottom right}), and these group galaxies in their real-space positions
({\it top right}).  In each case, galaxies in halos or groups with $N>4$ are
shown as colored points, with each halo or group represented by a unique color.
Large open circles are centered on the halo or group centers and have radii
equal to the halo virial radii ({\it left panels}) and the estimated group
virial radii ({\it right panels}).
}
\label{fig:nbodyslice}
\end{figure*}
%-----------------------------------------------------------------------

At low $N$, the multiplicity function is highly biased, both in shape and amplitude.
The abundance of groups relative to halos at a given multiplicity $N$ decreases when 
FoF splits these halos into smaller groups or merges them to form larger groups.
This decrease is countered by an increase due to the merging of smaller halos or the
splitting of larger halos.  The balance between these competing effects determines
whether the multiplicity function is biased or not.  For linking lengths near $b=0.2$,
merging dominates over splitting, which means that group abundances at a given 
multiplicity are mainly determined by a balance between halos at that $N$ merging to
yield larger groups and smaller halos merging to replenish the lost groups.  However,
this balance breaks at $N=1$ because, while FoF merges $N=1$ halos (i.e., isolated 
galaxies) to form larger groups, there are no smaller halos that can merge to 
replenish $N=1$ groups.  The abundance of $N=1$ groups is therefore necessarily
less than that of $N=1$ halos (it can only be more if the linking length is so small 
- approximately $b\sim0.1$ - that single galaxy groups splinter off in large numbers 
from larger halos).  Since most galaxies live in $N=1$ halos ($\sim70\%$ in these mock 
catalogs), merging a small fraction of them to form larger groups will fractionally 
increase the abundance of larger $N=2, 3, 4$, etc. groups significantly.  This is seen 
in Figure~\ref{fig:nbodymultbxy}: the abundance of $N=1$ groups is lower than that of
halos by $\sim20\%$ for $b=0.2$, causing the abundance of $N=2$ and $N=3$ groups to 
be $\sim50\%$ higher.  Only for $N>10$ does the group abundance settle down and become
unbiased.  This behavior is a fundamental limitation of the FoF algorithm, and it
has the consequence that group abundances can only be trusted for large multiplicity 
groups.

In redshift space, group finding is much more challenging because finger-of-god
distortions stretch groups along the line-of-sight, making it more likely that
single halos will be split into multiple groups and that neighboring halos will
be merged into the same groups.  Figure~\ref{fig:nbodyslice} illustrates these
effects by showing the performance of FoF in a small slice through a single mock
catalog (one HOD realization of the \texttt{LANL4.Mr20} mock catalog).  The top-left 
panel shows the mock galaxies in real space, with each $N>4$ halo denoted by a unique 
color.  The bottom-left panel shows the same galaxies in redshift space, where the 
line-of-sight is oriented along the $z$-axis of the mock cube.  Large open circles have
radii equal to the halo virial radii and are centered at the halo centers in real space,
and the galaxy centroids in redshift space.  We run our adopted FoF group-finder
(described in \S~\ref{groupfinder}) on the redshift-space mock and denote each
resulting $N>4$ group with a unique color in the bottom-right panel.  Finally, we
show the group galaxies' real-space positions in the top-right panel.  Large dotted 
circles are centered at the group centroids and have virial radii that are estimated
by assuming a halo mass function and a monotonic relation between group multiplicity 
and mass.  A visual comparison of the real- and redshift-space panels reveals many of 
the failure modes of FoF group-finding in redshift space.  The halo denoted by green 
in the left-side panels is fairly well recovered by FoF as the group denoted by green in
the right-side panels.  However, a couple of halo galaxies are missed in group finding, 
such as the one whose velocity moved it the furthest away from the center of the halo.
Most of the galaxies in the halo denoted by blue are linked together in the same group,
also denoted by blue.  However, many galaxies that do not belong to the ``blue'' halo
are also linked to the same group.  This is seen clearly in the top-right panel, where
seven of the ``blue'' group galaxies' real-space positions place them well outside the
halo.  A similar thing occurs to the halos and corresponding groups denoted by magenta 
and cyan.  Most of the galaxies in the large ``red'' halo are recovered correctly into
the ``red'' group, but there are some galaxies added to this group that do not belong
to the ``red'' halo, as well as a few galaxies that do belong to that halo, but have
splintered off into a different group (denoted by dark green).  Despite these 
imperfections, there is clearly a substantial correspondence between the groups 
identified by FoF and the true population of halos in this slice.

%-----------------------------------------------------------------------
\begin{figure}[t]
\epsscale{1.0}
\plotone{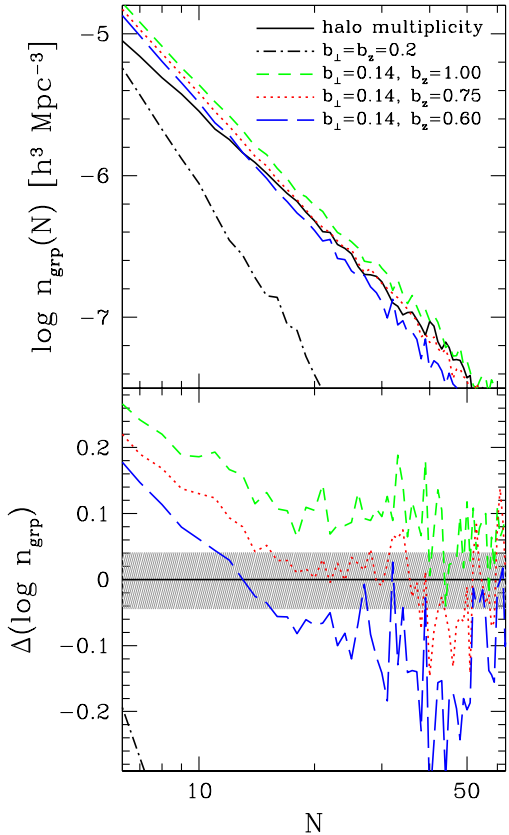}
\caption{
Same as Fig.~\ref{fig:nbodymultbxy}, but for redshift space.  The solid black
curve shows the input halo multiplicity function.  The dot-dashed black curve
shows the recovered group multiplicity function if a single linking length is used.
The other three curves show the recovered multiplicity functions for fixed
perpendicular and three different line-of-sight linking lengths, which are listed
in the top panel in units of the mean inter-galaxy separation.  All other
features are as in Fig.~\ref{fig:nbodymultbxy}.
}
\label{fig:nbodymultbxybz}
\end{figure}
%-----------------------------------------------------------------------

We now examine the relative multiplicity functions of groups and halos when the groups 
are identified in redshift space.  If we use the same linking length in transverse and 
line-of-sight directions, finger-of-god distortions will cause halos to be split into 
multiple small groups along the line-of-sight.  This is demonstrated by the dashed 
curve in Figure~\ref{fig:nbodymultbxybz}, which shows the multiplicity function of groups
identified with a single linking length of $b=0.2$.  The abundance of groups is
vastly underestimated for $N\gtrsim 5$, and the effect grows with $N$ because richer 
halos have higher velocity dispersions.  We therefore need to use different linking 
lengths in the line-of-sight and perpendicular directions.  We apply FoF to our 
redshift-space cube mocks for a grid of perpendicular and line-of-sight linking lengths 
and find that we can recover an unbiased multiplicity function at large $N$ for the right
combinations of linking lengths.  Figure~\ref{fig:nbodymultbxybz} shows one such
combination ($\bperp=0.14$, $b_z=0.75$) and demonstrates how the group multiplicity 
function changes with the line-of-sight linking length $b_z$.  Generally,
larger linking lengths in either direction lead to a higher abundance of groups at
large $N$.  We record all linking length combinations that yield unbiased multiplicity 
functions in the large $N$ regime and show the successful parameter space in
Figure~\ref{fig:linkinglengths.hod20}, as discussed in \S~\ref{groupfinder}.

%-----------------------------------------------------------------------
\begin{figure}[t]
\epsscale{1.0}
\plotone{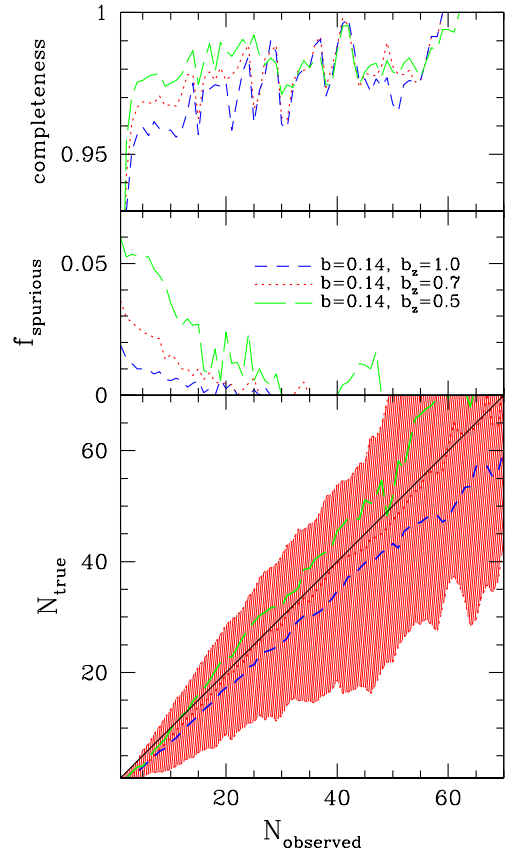}
\caption{
Effect of changing the FoF linking lengths on the relation between the distributions
of input halo richness and recovered group richness in redshift space, measured using
mock galaxy catalogs.  Each input halo is matched one-to-one to a recovered group
whenever possible; however, some halos have no corresponding group and some groups
have no one-to-one parent halo.  The top panel shows the halo completeness as a
function of halo richness, i.e., the fraction of halos at each richness that can be
matched one-to-one with a recovered group.  The middle panel shows the spurious
fraction of groups as a function of group richness, i.e., the fraction of groups
at each richness that cannot be matched one-to-one with a parent halo.
The bottom panel shows the relation between halo and group richness for halos and
groups that are matched one-to-one.  Middle curves show the median relation
and outer curves show the 10 and 90 percentiles (they enclose $80\%$ of the
group-halo pairs).  The area between these outer curves is shaded. In all panels,
different line types and colors show fixed perpendicular and different line-of-sight
linking lengths, which are listed in the top panel in units of the mean inter-galaxy
separation.  To avoid confusion, the 10 and 90 percentile curves (as well
as the shading between them) in the bottom panel are only shown for one of the
linking length combinations.  All results are averaged over twenty mock galaxy catalogs.
}
\label{fig:nbodycompbxybz}
\end{figure}
%-----------------------------------------------------------------------

Recovering an unbiased multiplicity function does not guarantee that the one-to-one 
relation between the multiplicities of halos and their recovered groups is also 
unbiased.  We therefore also investigate this relation.  As described in
\S~\ref{groupfinder}, we associate each halo to the recovered group that contains
the halo's central galaxy.  Groups that contain central galaxies from more than one
halo are associated with the halo with which they share the largest number of galaxies.
Halos that end up not being associated with any group are considered ``undetected,''
and groups that are not associated with any halo (i.e., they contain no halo central 
galaxies) are considered ``spurious''.  Once we have associated mock groups one-to-one
with their parent halos, we can look at the relation between the halo and group 
multiplicities (i.e., $\Ntrue$ vs. $\Nobs$).  In addition, we can look at the fraction
of halos that are detected and the fraction of groups that are spurious.
Figure~\ref{fig:nbodycompbxybz} shows how these relations depend on the line-of-sight
linking length.  The bottom panel of the figure shows one set of linking lengths 
($\bperp=0.14$, $b_z=0.70$) that yields an unbiased median relation between $\Ntrue$ 
and $\Nobs$, but the scatter around this relation is large and quite asymmetric.  
90\% of groups at a given $\Nobs$ are associated with halos that have up to 40\% 
higher and 60\% lower $\Ntrue$.  Increasing the line-of-sight linking length causes 
groups to grow and thus biases the median $\Ntrue$ vs. $\Nobs$ relation by tilting it 
toward larger $\Nobs$.  As before, we record all linking length combinations that 
yield unbiased median relations between group and halo multiplicities, and we show the 
successful parameter space in Figure~\ref{fig:linkinglengths.hod20}.

%-----------------------------------------------------------------------
\begin{figure}[t]
\epsscale{1.0}
\plotone{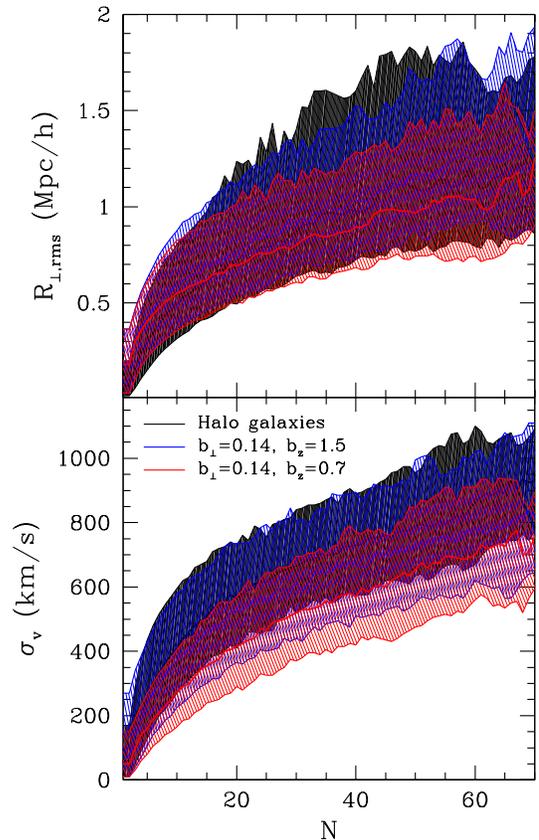}
\caption{
Effect of changing the FoF linking lengths on the size distribution of groups
in redshift space, measured using mock galaxy catalogs.  The top panel shows the 
projected 2-dimensional rms group radius distribution as a function of group
richness $N$.  The bottom panel shows the same for the 1-dimensional line-of-sight
velocity dispersion $\sigma_v$.  In both panels, the black curves and shading show 
the size distributions of galaxy systems that occupy the same dark matter halo and thus 
represent the ``true'' cases.  The sets of colored curves and shadings show the
size distributions of recovered groups for fixed perpendicular and three different 
line-of-sight linking lengths, which are listed in the bottom panel in units of the 
mean inter-galaxy separation.  Middle curves show the median relation and outer 
curves show the 10 and 90 percentiles.  The area between these outer curves is shaded.
All results are averaged over twenty mock galaxy catalogs.
}
\label{fig:nbodygrpstatsbxybz}
\end{figure}
%-----------------------------------------------------------------------

The top panel of Figure~\ref{fig:nbodycompbxybz} shows the completeness (fraction of 
halos that are associated one-to-one with groups) as a function of halo multiplicity
$\Ntrue$, and the middle panel shows the spurious group fraction as a function of
group multiplicity $\Nobs$.  Over a wide range of FoF linking lengths, the completeness 
for halos with $N\gtrsim 5$ is over 95\%, and the spurious fraction for groups with 
$N\gtrsim 5$ is less than 5\%.  Increasing the line-of-sight linking length causes
a drop in the halo completeness and a corresponding drop in the spurious group fraction,
since more halos get linked to the same groups.  For the final linking lengths that
we use (see \S~\ref{groupfinder}), the halo completeness is greater than 97\% and
the spurious group fraction less than 1\% for $N\gtrsim 10$.  The high completeness
and low spurious fraction are a result of how we associate groups to halos.  Since we
only require a group to have a halo's central galaxy in order to be associated with it,
most groups and halos have one-to-one associations.  If we used a more stringent 
criterion for group-halo association, for example by requiring that a group contain 
some minimum fraction of a halo's galaxies, then the halo completeness would be lower 
and the spurious group fraction higher, but the scatter in $\Ntrue$ vs. $\Nobs$ would 
be reduced.  The three panels of Figure~\ref{fig:nbodycompbxybz}, put together, 
characterize the errors in the FoF group finder.  Changing the definition for how groups 
are associated to halos does not change the errors in group-finding; it merely 
redistributes the errors among the three panels.

In addition to requiring that our groups have unbiased abundances and multiplicities,
we would also like them to have unbiased size distributions.  For every group in our 
redshift-space cube mocks, we measure the projected rms radius and the line-of-sight 
velocity dispersion of galaxies.  We compare these to the projected rms radii and
actual velocity dispersions of halo galaxies.  Figure~\ref{fig:nbodygrpstatsbxybz}
shows the median, 10th, and 90th percentile projected size and velocity dispersion as a 
function of multiplicity for halos, compared to that for groups identified with two 
different line-of-sight linking lengths.  Increasing the line-of-sight linking length
produces groups with higher velocity dispersions, but it has less impact on 
the projected size distributions.  The opposite is naturally true when we increase the
perpendicular linking length.  Linking length combinations that yield groups with
unbiased abundances and projected sizes tend to yield velocity dispersions that are 
biased low.  This is illustrated in Figure~\ref{fig:nbodygrpstatsbxybz}, which
shows that the linking length combination $\bperp=0.14$, $b_z=0.7$ yields groups with
velocity dispersions that are $\sim 20\%$ too low relative to halos.  The line-of-sight
linking length must be more than doubled to repair this bias, but then the abundances
of groups would be too high.

%-----------------------------------------------------------------------
\begin{figure}[t]
\epsscale{1.0}
\plotone{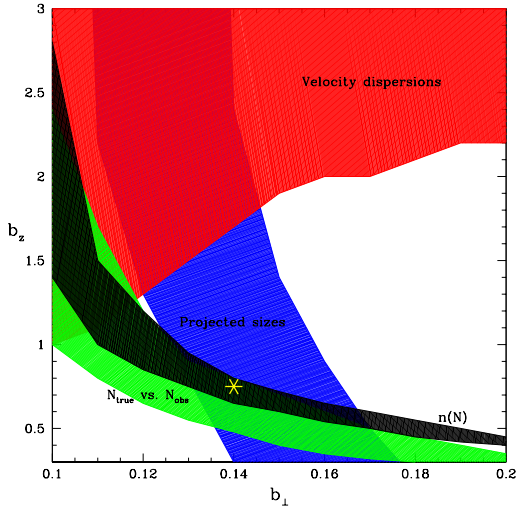}
\caption{
Same as Fig.~\ref{fig:linkinglengths.hod20}, but using the \texttt{.Mr20b} set of 
mock catalogs, which are constructed with a different input relation between halo 
richness and dark matter halo mass, as described in \S~\ref{mocks}.
}
\label{fig:linkinglengths.hod20b}
\end{figure}
%-----------------------------------------------------------------------
%-----------------------------------------------------------------------
\begin{figure}[t]
\epsscale{1.0}
\plotone{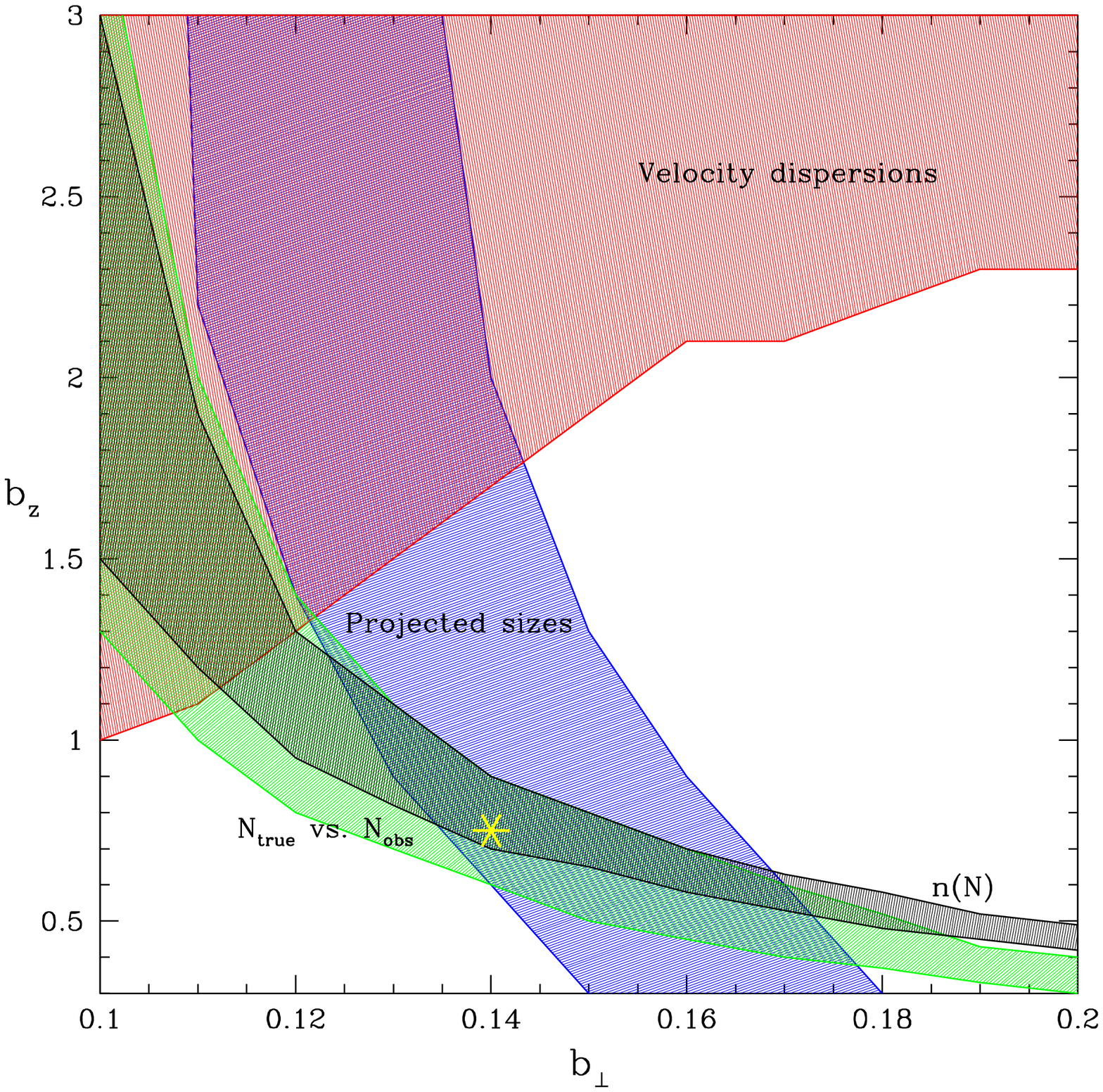}
\caption{
Same as Fig.~\ref{fig:linkinglengths.hod20}, but using the \texttt{.Mr19} set of 
mock catalogs, described in \S~\ref{mocks}.
}
\label{fig:linkinglengths.hod19}
\end{figure}
%-----------------------------------------------------------------------
%-----------------------------------------------------------------------
\begin{figure}[t]
\epsscale{1.0}
\plotone{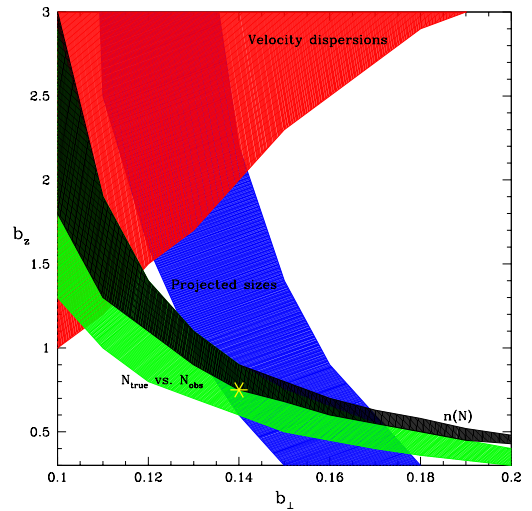}
\caption{
Same as Fig.~\ref{fig:linkinglengths.hod20}, but using the \texttt{.Mr18} set of 
mock catalogs, described in \S~\ref{mocks}.
}
\label{fig:linkinglengths.hod18}
\end{figure}
%-----------------------------------------------------------------------

Figure~\ref{fig:linkinglengths.hod20} shows the linking length parameter space that
satisfies each of the above tests.  As discussed in \S~\ref{groupfinder}, there
is no combination of perpendicular and line-of-sight linking lengths that yields
groups with unbiased abundances, projected sizes, and velocity dispersions, even at
high multiplicity.  We choose to sacrifice velocity dispersions and adopt the
parameters $\bperp=0.14$, $b_z=0.75$.  All the above tests and resulting choice
of linking lengths were done using the \texttt{.Mr20} mock catalogs.  Since we plan
to use our group catalog to constrain the HOD, it is vital that our choice of
linking lengths does not depend sensitively on the input HOD assumed when constructing
the mocks.  For this reason, we repeat all the above tests with the \texttt{.Mr20b}
mock catalogs, which use a different input HOD to model the same $Mr20$ sample of 
SDSS galaxies.  The results are shown in Figure~\ref{fig:linkinglengths.hod20b}.
It is clear that our adopted group finder performs equally well in both sets of mock 
catalogs, demonstrating that our choice of linking lengths is insensitive to the
underlying HOD.
It is also important to show how well our linking lengths work on lower luminosity
galaxy samples, since we apply them to the SDSS $Mr19$ and $Mr18$ samples.  We thus
repeat our mock tests with the \texttt{.Mr19} and \texttt{.Mr18} mock catalogs and show
the results in Figures~\ref{fig:linkinglengths.hod19} and~\ref{fig:linkinglengths.hod18},
respectively.  The figures show that lower luminosity (higher density) samples
require slightly higher line-of-sight linking lengths in order to retain unbiased
multiplicity functions.  However, this effect is small.  When applied to the 
\texttt{.Mr18} mock catalogs, our adopted linking lengths yield a multiplicity function 
that is 10\% too low in amplitude.  Overall, Figures~\ref{fig:linkinglengths.hod20},
\ref{fig:linkinglengths.hod20b}, \ref{fig:linkinglengths.hod19}, 
and~\ref{fig:linkinglengths.hod18} demonstrate that our choice of linking lengths
is fairly robust.

%#############################################################################

\def\baselinestretch{1}

\bibliographystyle{apj}
\bibliography{}

%-----------------------------------------------------------------------

\end{document}